\documentclass[a4paper,10pt]{article}
\usepackage[centertags]{amsmath}
\usepackage{amssymb}
\usepackage[T1]{fontenc}
\usepackage{bera}
\usepackage[margin=1.0in]{geometry}
\usepackage{url}
\usepackage{hyperref}
\usepackage{enumerate}
\usepackage{epsfig}
\usepackage[square, comma, numbers, sort&compress]{natbib}
\usepackage{soul}

\begin{document}
\title{Dynamics of quantum quenching for BCS-BEC systems in the shallow BEC regime}
\author{\normalsize{Analabha Roy} \\
\normalsize{S.N Bose National Centre for Basic Sciences}\\
\normalsize{Sector 3, Block JD, Salt Lake, Kolkata 700098, India.}\\
\normalsize{Email:\url{daneel@bose.res.in}}
}
\maketitle
\begin{abstract}
The problem of coupled Fermi-Bose mixtures of an ultracold gas near a narrow Feshbach resonance is approached through the time-dependent and complex Ginzburg-Landau (TDGL) theory. The dynamical system is constructed using Ginzburg-Landau-Abrikosov-Gor'kov (GLAG) path integral methods with the single mode approximation for the composite Bosons.  {The equilibrium states are obtained in the BEC regime for adiabatic variations of the Feshbach detuning along the stationary solutions of the dynamical system}. Investigations into the rich superfluid dynamics of this system in the shallow BEC regime  {yield} the onset of multiple interference patterns in the dynamics as the system is quenched from the deep-BEC regime. This results in a partial collapse and revival of the coherent matter wave field of the BEC, whose temporal profile is reported.
\end{abstract}

\section{Introduction}
\label{sec:intro}
Over the course of the last two decades, there has been a groundswell of theoretical and experimental interest in ultracold gases of alkali metals confined in optical and/or magnetic traps. These systems of ultracold gases have proved to be extremely robust and tunable systems for studying condensed matter physics in regimes that are inaccessible in solid state systems~\cite{lewenstein:review}. During this time, physicists also became specifically interested in condensates of Fermionic alkali atoms (such as $^{6}Li$), and obtaining a \textit{BCS superfluid} of Cooper pairs similar to those seen in solid state superconductors (BCS theory). If the effective attraction between Cooper pairs can be rendered sufficiently strong, Cooper pairs of Fermions are no longer merely correlated and far apart as in traditional BCS systems, but have much smaller correlation lengths approaching the interparticle spacing. Thus, they can be treated as \textit{composite Bosons}, causing the system to undergo a \textit{BCS-BEC crossover} to a BEC superfluid. 

Theoretical work on the single channel model of the BCS-BEC crossover over the course of the late $1980$s and $1990$s~\cite{springerlink:10.1007/BF00683774,randeria:fermigas,randeria:bcsbec3,randeria:bcsbec2d,dreschler:bcsbec} motivated numerous experiments with laser cooling and trapping of Fermions, culminating in the observation of this crossover in the early $2000$s~\cite{zwierlein:bcsbecexpt,thomas:superfluid,fermionic:cond:K, collective1}. In the late $1990$s, the more informative \textit{dual-channel} model of \textit{resonant superfluidity} was proposed, by Holland \textit{et al}~\cite{holland:bcsbec}, and by Timmermans \textit{et al}~\cite{timmermans}. This model, built from the \textit{Timmermans' Hamiltonian}, is a generalization of the Dicke (Tavis - Cummings) model in quantum optics~\cite{yuzbashyan}. The attractive interaction between the Fermions can be controlled by tuning a homogeneous magnetic field to their \textit{Feshbach resonances}~\cite{timmermans}, which are caused in a two-particle system  by coupling bound states in a close channel with states in the scattering continuum. The effective scattering length can be tuned simply by varying the external magnetic field that controls the net magnetic moments of the different channels. In this dual channel model, sufficiently strong resonances cause bound states to form in the closed channel. Thus, Cooper pairs of Fermions can \textit{physically combine} into Bosonic molecules~\cite{bcsbec:ohashi}, entities referred to variously as \textit{composite Bosons}, \textit{diatomic molecules}, or \textit{quasimolecular states} in the literature. These composite Bosons have extremely long lifetimes near a Feshbach resonance~\cite{cubizolles:feshbachlife1,strecker:feshbachlife2}, and repel each other~\cite{petrov:repulsivebosons}, facilitating their condensation to a \textit{BEC superfluid}~\cite{donley:becquench,bourdel}. This area of research has gained enormous interest over the last two decades. An overview of the phase portrait of a population balanced Fermi gas can be found in~\cite{randeria:bcsbecnature}, and reviews of the current status of research can be found in~\cite{stringari:review,bloch:review}. 

The possibility of observing the nonequilibrium dynamics of coherent quantum states via \textit{quenching} (diabatic variations of the system parameters) is one that is unique to systems of ultracold gases, and unavailable in other similar condensed matter systems. Post-quenched dynamics in cold atom systems have been studied and reported. In particular, the \textit{collapse and revival} of a post-quenched coherent BEC state in an optical lattice~\cite{greiner:colrev} generated considerable interest. Quenched dynamics have also been investigated for Fermi-Bose mixtures, experimentally~\cite{olsen:rabifermi}, as well as theoretically~\cite{andreev:noneqmbcsbec,barankov:bcsbecbloch,uys:hyperbolicbcsbecdyn,jackpu:bcsbecdyn,yuzbashyan,yuzbashyan2}. The possibility of observing a \textit{collapse and revival} phenomenon as the system is quenched past the BCS-BEC crossover has been raised~\cite{huang:becbcs2}. More recently, nonequilibrium dynamics in condensed matter physics have also been described by the complex Ginzburg-Landau equation~\cite{ginzburglandau:review}. The nonlinear damping in the dynamics produces very rich and diverse behavior in Fermi-Bose mixtures~\cite{machida:dynamics}. The mean field dynamics of BCS superconductivity have been obtained from microscopic models via Ginzburg-Landau-Abrikosov-Gor'kov (GLAG) theory. A similar approach has proven successful for BCS-BEC systems as well, thus leading to their description by the TDGL equation for the single channel case~\cite{randeria:bcsbec3}. More recently, the applicability of TDGL dynamics have been demonstrated in the dual channel case~\cite{machida:dynamics}. This motivates the  use of this treatment to study the dynamics of coherent matter waves in BCS-BEC systems in this report.

This paper focuses on the dynamics of the TDGL equation in coupled BCS-BEC systems, and the dynamics of quantum quenching therein. The most general state of this system across the phase diagram is that of a \textit{Fermi-Bose mixture}, where the composite Bosons coexist with correlated Fermions, and their dynamics are linked in the mean field by the two-channel scattering process through coupled Ginzburg-Landau and Gross-Pitaevski equations. The mixture is characterized by two distinct phases, the Fermi (BCS) superfluid phase given by the order parameter $\Psi_1$, and the Bose (BEC) superfluid phase given by the order parameter $\Psi_2$. The Fermi superfluid consists of distinct Cooper pairs, and the Bose superfluid consists of quasimolecular Bosons. This 2-channel model is more faithful to the microscopic nature of the system than the single-channel model, especially when Feshbach resonances are involved. The dynamics of the coupled BCS-BEC phases is richer than that described by a single-phase model in the single channel case, since the Ginzburg-Landau dynamics of the latter is only the damped one except in the BEC regime. This is especially true if the Feshbach resonance is narrow, which is the case being discussed in this work.  {Though many current experiments correspond to broad resonances, the consideration of a narrow Feshbach resonance here is not unrealistic at all}~\cite{rembert:stoof,stoof:other}. The paper reconstructs the time dependent Ginzburg Landau equations for the dynamics of a Fermi-Bose mixture in the two channel case starting from the many-body functional field integral for the Timmermans' Hamiltonian. This Hamiltonian describes the Fermions by the BCS Hamiltonian and the composite Bosons by the Bose-Hubbard Hamiltonian~\cite{timmermans}, and includes a resonant coupling between the two species. The path integral is written as a functional of the superfluid gap parameter $\Delta$ and the boson coherent state (all of them are taken to lie in the zero momentum state) $b_0$, taken to be a c-number. The mean-field dynamics is subsequently derived, and the equilibrium densities evaluated at the stationary solutions. The paper then details investigations of the dynamics of quantum quenching, and reports the possibility of collapse and revival in the full matter wave of the BEC at large times.  Section~\ref{sec:dyneqns} begins by outlying the formalism that obtains the TDGL dynamics from the Timmermans' Hamiltonian. Section~\ref{sec:fpchempot} reports the study of the stationary solutions of the TDGL dynamical system in 3 dimensions. The chemical potential and condensate fractions are evaluated as the Feshbach detuning is varied adiabatically. Section~\ref{sec:quench} looks at the dynamical evolution of the BEC (which can be seen directly in the lab via time-of-flight absorption) as the Feshbach detuning is varied \textit{diabatically} (ie 'quenched') to the shallow BEC regime. Concluding remarks are made in Section~\ref{sec:conclusion}.
\section{Dynamical Equations of Motion}
\label{sec:dyneqns}
The treatment that obtains the coupled TDGL dynamics of this system closely follows that which obtains the conventional TDGL dynamics of the single channel model, as applied by Huang, Yu and Yin~\cite{huang:bcsbecgp}. The formalism is applied to the dual channel case in a manner similar to the treatments by Machida and Koyama~\cite{machida:dynamics}, except for the nature of the Bosonic states, which is approximated by a single mode. It is assumed that, at $T=0$, the time-scales of the dynamics are sufficiently weak so as to not induce Boson excitations above the ground state, an assumption justified in the context of dynamics in greater detail in the literature~\cite{andreev:noneqmbcsbec}. This approximation greatly simplifies the dynamics by removing any spatial information in the composite Boson amplitudes from the beginning. This approximation also allows for rapid transitions near unitarity.

The $D-$dimensional Fermion and zero momentum composite Boson fields in this system are represented by the operators  $\phi_\sigma(x)$ and $b_0$  respectively, with the index $\sigma = \uparrow,\downarrow$ representing the Fermion pseudospin. The dual-channel \textit{Timmermans' Hamiltonian}~\cite{timmermans} $H_{tm}$ for a Fermi-Bose mixture at $T=0$ for a unit volume is 
\begin{equation}
 H_{tm} = \int{{\mathrm d}^Dx} \times {\mathcal H_{tm}} (x),
\end{equation}
where
\begin{multline}
\label{timmermans}
{\mathcal H_{tm}}(x) =  \sum_{\sigma} \bigg\{ \phi^\dagger_\sigma(x) \left[h({\bf r})-\mu_F\right]\phi_\sigma(x) \bigg \} - |u_F| \phi^\dagger_\uparrow(x)\phi^\dagger_\downarrow(x)\phi^{}_\downarrow(x)\phi^{}_\uparrow(x)\\ 
+\left[ 2\nu-\mu_B \right]b^\dagger_0(t')b^{ }_0(t')+ u_B b^\dagger_0b^{ }_0\left(b^\dagger_0b^{ }_0-1 \right)\\
+g_r\left[b^\dagger_0(t') \phi_\uparrow(x) \phi_\downarrow(x) + h.c\right].
\end{multline}
Here, $x=({\bf r},t')$. Equation~\ref{timmermans} describes a system of ultracold electrically neutral two-component Fermions interacting attractively. The first line in equation~\ref{timmermans} represents the \textit{Fermi-BCS} part of the Hamiltonian. Here, $h({\bf r})=\left[ - \frac{\nabla^2}{2m}+V({\bf r}) \right]$ is the single particle Hamiltonian, and $m$ is the Fermion mass (the mass of the composite Bosons is thus $2m$). The second line represents the Hamiltonian of the composite Bosons~\cite{timmermans, huang:becbcs2}. Here, the Feshbach threshold energy (also called the Feshbach 'detuning' from the molecular channel to the continuum~\cite{timmermans}) is represented by $2\nu$, and $u_B$ represents the amplitude of the repulsion between the composite Bosons. The final line describes the Feshbach resonance that leads to the Fermions binding to (or dissociating from) the composite Bosons, with  $g_r$ representing the atom-molecule coupling. Note that the chemical potentials satisfy $\mu_B=2\mu_F$. The path integral grand partition function $Z$ is defined by
\begin{equation}
\label{pathintegral}
 Z=\int{{\mathrm D}[\bar{\phi},\phi]{\mathrm D}[b^*,b]}e^{-S_{\phi b^{ }_0}},
\end{equation}
where ${\mathrm D}[\bar{\phi},\phi]$ and ${\mathrm D}[b^*,b]$ are the path integral measures of the Fermion and Boson fields respectively, and $t'$ is the imaginary time. The action $S_{\phi b^{ }_0}$ is given by
\begin{multline}
\label{pathintegral:action}
 S_{\phi b^{ }_0} = \sum_{\sigma} \int{{\mathrm d}^Dx} \int^\infty_0{{\mathrm d}t'} \bigg[ \bar{\phi}_{\sigma}(x)\partial_{t'} \phi_{\sigma}(x)+ b^*_0(t')\partial_{t'}b^{ }_0(t') + 
{\mathcal H_{tm}}(\bar{\phi},\phi,b^*_0,b^{ }_0)\bigg].
\end{multline}
This integral can be evaluated by introducing the macroscopic gap parameter $\Delta(t')$ into equation~\ref{pathintegral} via the Gaussian identity
\begin{equation}
\int {\mathcal D}[\Delta^*,\Delta] \exp{\left[-\int^\infty_0{{\mathrm d}t'}\frac{\Delta^*(t')\Delta(t')}{|u_F|}\right]}=1,
\end{equation}
where the Fermion field (characterized by the gap parameter) is spatially homogeneous due to it's coupling  to the spatially homogeneous Bose field $b_0$ by momentum conservation. Performing the Hubbard-Stratonovich transformation, $\Delta\rightarrow \Delta + |u_F|\phi_\downarrow\phi_\uparrow $ and $\Delta^*\rightarrow \Delta^* + |u_F|\bar{\phi}_\uparrow\bar{\phi}_\downarrow$, cancels out the four-Fermion term from $H_{tm}$ in equation~\ref{pathintegral:action}. Now, integrating out the Fermion fields $\phi,\bar{\phi}$ remaining in equation~\ref{pathintegral} using formal Grassman calculus yields
\begin{equation}
 Z = \int{{\mathrm D}[b^*,b] {\mathrm D}[\Delta^*,\Delta]} e^{-S_{\Delta b^{ }_0}},
\end{equation}
where
\begin{multline}
\nonumber
 S_{\Delta b^{ }_0} = \int{{\mathrm d}t'} \bigg\{ \left[ 2\nu-\mu_B \right] |b^{ }_0(t')|^2 + u_B |b_0(t')|^2\left[|b_0(t')|^2-1 \right] + b^*_0(t')\partial_{t'}b^{ }_0(t')+ \frac{|\Delta(t')|^2}{|u_F|} \bigg\} + \\
\int{{\mathrm d}^Dx \times {\mathrm d}t'} \ln\det{\bf M}(x).
\end{multline}
Here,
\begin{equation}
  {\bf M}(x) \equiv \left[ \begin{array}{cc}
							    \partial_{t'}-\mu_F+h({\bf r})  & \Delta(t')+g_r b^{ }_0(t') \\
							    \Delta^*(t')+g_r b^*_0(t')  & \partial_{t'}+\mu_F-h({\bf r})
                                                             \end{array} \right]. 
\end{equation}
The action $S_{\Delta b^{ }_0}$ is split into a field independent part $S_0= \int{{\mathrm d}^Dx {\mathrm d}t'} \ln\det{M_0(x)}$, where
\begin{equation} 
{\bf M}_0(x)\equiv\left[ \begin{array}{cc}\partial_{t'}-\mu_F+h({\bf r})&0\\
0&\partial_{t'}+\mu_F-{h}({\bf r})\end{array}\right], 
\end{equation}
and a field dependent part $S_{eff}$ which vanishes when $\Delta$ and $b_0$ do. Expanding $S_{eff}$ to the fourth order in
$\Delta + g_rb_0$~\cite{huang:bcsbecgp} and performing gradient expansion results in
\begin{multline}
\label{gdexp}
S_{eff}\approx\int{d}^{\mathcal{D}}x\bigg\{\,d\,\left[\Delta^*(t')+g_rb^*_0(t')\right]\partial_{t'}\left[ \Delta(t') + g_rb^{ }_0(t') \right]+
 \frac{|\Delta(t')|^2}{|u_F|}\\
-a|\Delta(t')+g_rb^{ }_0(t')|^2+ \frac{1}{2}b|\Delta(t')+g_rb^{ }_0(t')|^4+ \\ 
\left[2\nu-\mu_B\right]|b_0(t')|^2  + u_B|b_0(t')|^2\left[|b_0(t')|^2-1 \right]  +b^*_0(t')\partial_{t'}b^{ }_0(t')\bigg\}.
\end{multline}
Here,
\begin{eqnarray}
\label{abc}
a&=& \int{\mathrm d}^{\mathcal{D}}x'\times Q(x-x'/2,x+x'/2),\nonumber\\
b&=&\int{\prod_{i=1}^D{\mathrm d}^{\mathcal{D}}x_i} \times R(x,x_1,x_2,x_3),
\end{eqnarray}
the coefficient $d$ is obtained from~\cite{machida:dynamics},
\begin{equation}
\label{d}
d=\lim_{\omega\rightarrow 0}\int{\mathrm d}^{\mathcal{D}}x'\times {e^{i\omega{t'}}-1\over i\omega} Q(x-x'/2,x+x'/2).
\end{equation} 
Here, ${\mathrm d}^{\mathcal{D}}x={\mathrm d}^Dx \times dt'$, where ${\mathrm d}^{D}x$ is the measure of integration over all $D$ spatial degrees of freedom, and 
\begin{eqnarray} 
Q(x_1,x_2) &=& G_+(x_1,x_2)G_-(x_2,x_1) \nonumber \\
R(x_1,...,x_4)&=&G_+(x_1,x_2)G_-(x_2,x_3)G_+(x_3,x_4)G_-(x_4,x_1),\nonumber \\
\end{eqnarray}
with $x'=({\bf r}',t')$, and the Gor'kov Green's function, $ {\bf G}(x)= {\bf M}^{-1}_0(x)=\bigg[\begin{array}{cc}G_+(x)&0\\
0&G_-(x)\end{array}\bigg]$, which is obtained from the Green's function of a noninteracting Fermi gas i.e.
\begin{equation}
\label{greens:function}
 \left[ \partial_{t'} \mp \mu \pm h({\bf r}) \right] G_\pm(x-x_1) = \delta^D({\bf r}-{\bf r}_1) \delta(t'-t'_1).
\end{equation}
Note that the product in the expression for $b$ above is over the spatial dimensions $D$ only, whereas the measure is over the space-time dimensions $\mathcal{D}$. The mean field equations of motion of the order parameters can be obtained by equating the functional derivatives $\delta S_{eff}/\delta\Delta^*(t')$ and   $\delta S_{eff}/{\delta b^*_0(t)}$ to $0$ after analytically continuing the time to the real axis by substituting $it \rightarrow t$. This yields the final dynamical equations for this system
\begin{eqnarray}
\label{dynamical_system}
\dot \Psi_1 + i\gamma \left( \Psi_1-\Psi_2 \right) - i\alpha \Psi_1 +i\beta |\Psi_1|^2 \Psi_1 &=& 0 \nonumber \\
\dot \Psi_2 +2 i \lambda \Psi_2 +2 i\chi|\Psi_2|^2\Psi_2 - i\kappa\gamma \left(\Psi_1-\Psi_2 \right) &=& 0.
\end{eqnarray}
Here,
\begin{eqnarray}
\label{vartrans}
&&{\Psi}_1 \equiv \frac{\Delta + g_rb_0}{|\mu_F|\sqrt{\mathcal N}}, \nonumber \\
&&{\Psi}_2 \equiv \frac{g_rb_0}{|\mu_F|\sqrt{\mathcal N}},
\end{eqnarray}
where $\mathcal{N}$ is the total (Fermion) particle number. The constants expressed as Greek letters in equation~\ref{dynamical_system} are given by
\begin{equation}
\label{greek_consts}
\begin{array}{ll}
 \alpha\equiv  a|\mu_F|,		&\lambda \equiv \bigg[\nu+|\mu_F|-\frac{u_B}{2} \bigg]|\mu_F|d,  \\
					&					  			 \\
\beta\equiv  b|\mu_F|^3{\mathcal N},    &\sigma \equiv   \frac{|\mu_F|}{g_r},			  	 \\
					&					  			 \\
\gamma\equiv  |\frac{\mu_F}{u_F}|, 	&\chi\equiv  d u_B|\mu_F|\sigma^2{\mathcal N}.			 \\
					&					  			 \\
\kappa  \equiv  g^2_rd,  		& 						  		 \\
\end{array}
\end{equation}
Finally, note that time has been rendered dimensionless via the transformation $t\rightarrow{1\over |\mu_F|d}\times t$, and the chemical potential is presumed to be negative. Thus, the dynamics of the Fermi-Bose mixture is that of a system where the Fermi and Bose fields evolve according to \textit{coupled} Ginzburg-Landau Gross-Pitaevski-Bogoliubov dynamics. The coupling is caused in the Ginzburg-Landau case by the order parameter $\Delta$ getting \textit{nonlinearly dressed} by the Bose field $g_rb_0$, and in the Gross-Pitaevski case by a harmonic coupling to $\Delta$.

It is also noted that the dynamical system bears a resemblance to that of a molecular BEC of atomic Bosons in a resonance effective field theory as proposed by Kokkelmans and Holland in 2002~\cite{kokkelmans:molbecdyn}. The role of the pairing field of noncondensed atoms (represented by the 'anomalous density' of noncondensed pairs) in that system is assumed by the Fermion gap parameter $\Delta$ (related to the anomalous Cooper pair density by $\Delta^* \sim \sum_{\bf k}\langle a_{{\bf k}\uparrow} a_{-{\bf k}\downarrow}\rangle$) in this one. This strengthens the analogy between Cooper pairs in Fermi-Bose systems and noncondensed atoms in Bose-Bose systems.

The dynamics is investigated for a system confined in a 3 dimensional box where the confinement $\mathcal{V}=\mathcal{L}^3$ is much larger than the inter-particle spacing, effectively treating the trap as homogeneous. The constants $a,b$ and $d$ can be evaluated from equations~\ref{abc} and~\ref{d} by using the Green's function for a free particle. This yields~\cite{machida:dynamics,huang:bcsbecgp}
\begin{equation} 
\label{Dl0_equation}
\begin{array}{lll}
a=\sum_{|{\bf k}|<k_R}\frac{1}{2\epsilon_{k}}, & b=\sum_{|{\bf k}|<k_R}\frac{1}{4\epsilon_{k}^3},& d=\sum_{|{\bf k}|<k_R}\frac{1}{4\epsilon^2_{\bf k}}.
\end{array}
\end{equation}
Here, $\epsilon_k=\frac{k^2}{2m}-\mu_F$ is the Fermion energy and $k_R={2\pi\over R}$ is the renormalization cutoff in momentum, placed to counter ultraviolet divergences. An important caveat here is that the integrals in equation~\ref{Dl0_equation} contain singularities if $\mu_F\geq\frac{k^2_R}{2m}$. Thus, this formalism breaks down in that regime, which corresponds to regions where the BCS state dominates over the BEC state, well beyond the current region of interest at a negative $\mu_F$. Going to the continuum limit by substituting for the formal sum $\sum \rightarrow \int\mathrm{d}^3k$ in equations~\ref{Dl0_equation} and performing the integrals results in
\begin{eqnarray}
\label{greek_consts_chempot}
\alpha\left(\epsilon_F\right)  &=& \frac{mk_R\epsilon_F}{2\pi^2} - \frac{\left(2m\epsilon_F\right)^{3/2}}{4\pi^2} \arctan{\frac{k_R}{\sqrt{2m\epsilon_F}}},\nonumber \\
\beta\left(\epsilon_F\right)   &=& \frac{\left( {2m\epsilon_F} \right)^{3/2}}{128\pi}\times\mathcal{N}, \nonumber \\
\gamma\left(\epsilon_F\right)  &=&  |\frac{\epsilon_F}{u_F}| =\frac{mk_R\epsilon_F}{2\pi^2}+\frac{\epsilon_F}{|u^0_F|} , \nonumber \\
\kappa\left(\epsilon_F\right)  &=&  \frac{g^2_r}{32\pi}\times \frac{\left(2m\epsilon_F\right)^{3/2}}{\epsilon_F^2},\nonumber \\
\lambda_\nu\left(\epsilon_F\right) &=&  \frac{\nu+\epsilon_F-\frac{1}{2}u_B}{32\pi}\times \frac{\left(2m\epsilon_F\right)^{3/2}}{\epsilon_F}, \nonumber \\
\sigma\left(\epsilon_F\right)  &=&   \frac{\epsilon_F}{g_r},\nonumber \\
\chi\left(\epsilon_F\right)    &=&  \frac{u_B\sigma^2\left(\epsilon_F\right)}{32\pi}\times \frac{\left(2m\epsilon_F\right)^{3/2}}{\epsilon_F}\times\mathcal{N}, 
\end{eqnarray}
where the dimensionless constants in Greek letters are now functions of the chemical potential $\epsilon_F\equiv|\mu_F|$\footnote{Note that, in general, $\epsilon_F$ as it is defined here does not equal to the Fermi energy.}, and the $\nu$-dependence on $\lambda_\nu$ has been emphasized by subscript. In the equations above, the relation
\begin{equation}
\frac{1}{|u_F|} = \frac{1}{|u^0_F|}+\sum_{|{\bf k}|<k_R}\frac{1}{2\epsilon^0_k},
\end{equation}
has been used to obtain the expression for $\gamma$. Here, $\epsilon^0_k = \frac{k^2}{2m}$, $u^0_F$ is the bare interaction $4\pi a_s/m\mathcal{V}$, and $a_s$ is the s-wave scattering length controlled by Feshbach resonance. This relation comes about as a consequence of renormalizing the BCS gap equation so as to counter ultraviolet divergences in the single channel model~\cite{randeria:fermigas,randeria:bcsbec2d,huang:bcsbecgp}.  { Finally, note that the inability to renormalize this theory analytically disallows taking the theoretical limit of $k_R\rightarrow\infty$ for some of the constants above. This problem will be discussed further in the subsequent section.}
\section{Stationary solutions and Chemical Potential}
\label{sec:fpchempot}
\begin{figure}[h!bt]
\ \epsfig{file=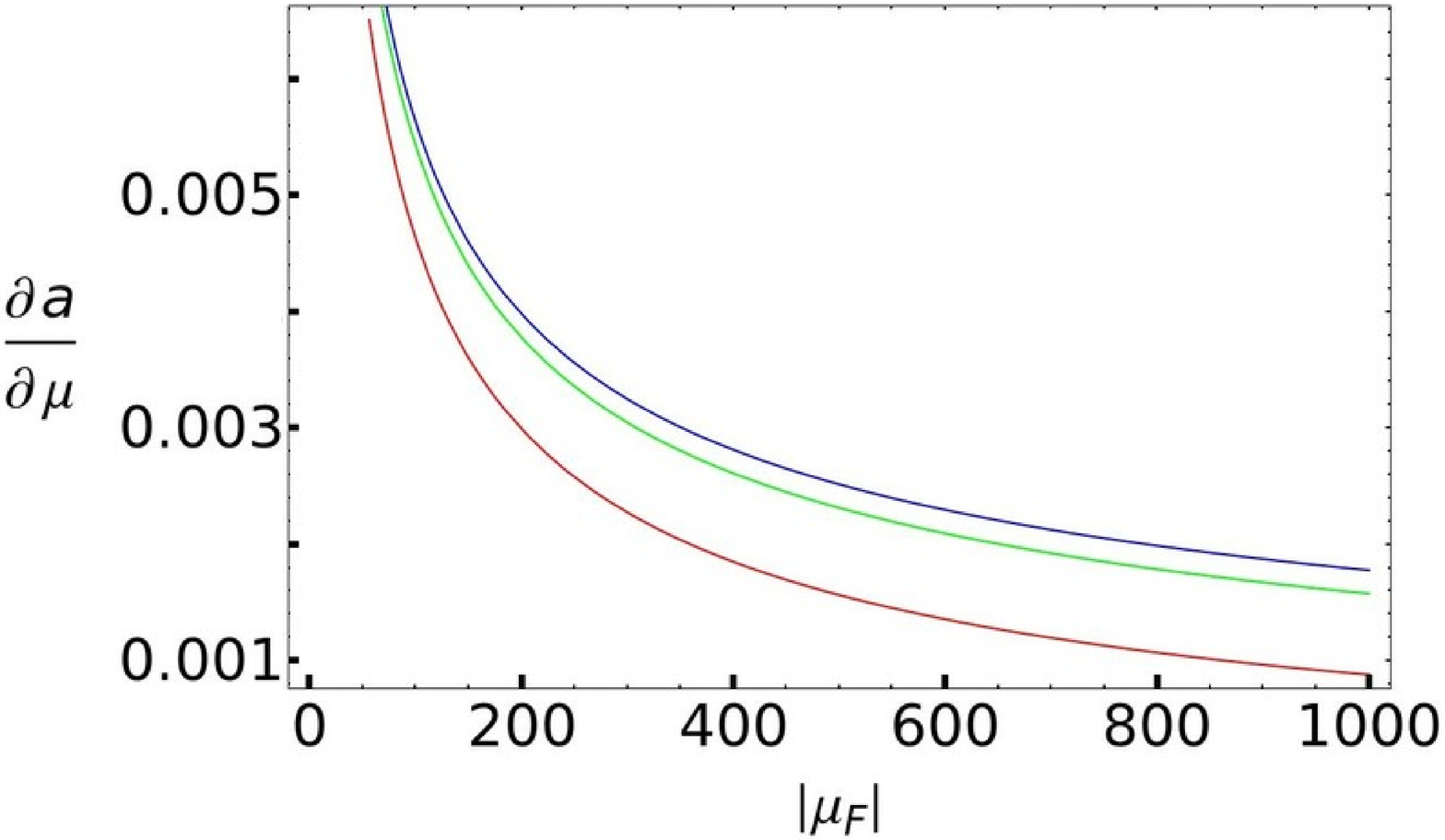,height=3.0in,width=4.5in}
\caption{(Color online) Plots of $\partial a/\partial\mu$ as a function of $|\mu_F|$ for $k_R=100$ (red), $500$ (green), and $10^9$ (blue). Here, $m=1$ and $\mathcal{V}=1$. Notice the rapid rate of convergence as $k_R\rightarrow\infty$.}
\label{fig:xi}
\end{figure}

For a complete phenomenological description of the dynamics, number conservation has to be satisfied and used as a constraint at $t=0$ to obtain the chemical potential $\epsilon_F=|\mu_F|$ where $\mu_F = {\mu_B \over 2}$ . In order to do this, the action $S_{eff}$ from equation~\ref{gdexp} is investigated at small temperature $\beta=1/k_BT$. Since we only consider regimes with negative chemical potential, all the Fermions in the gas are correlated and no 'free Fermions' remain~\footnote{The number density of uncorrelated Fermions, obtained from the 'free Fermion' term $S_0= \int{{\mathrm d}^Dx {\mathrm d}t'} \ln\det{M_0(x)}$ in section~\ref{sec:dyneqns}, is $n_{\bf k}=\Theta(\mu_F-\epsilon_{\bf k})$, which is $0$ if $\mu_F<0$.}. Thus, in the stationary case, 
\begin{equation}
\label{gdexp:hom}
S_{eff}(\beta)\approx\int^\beta_0{\mathrm d}t'\bigg\{\frac{|\Delta|^2}{|u_F|}-a|\Delta+g_rb^{ }_0|^2 + \left[2\nu-\mu_B\right]|b_0|^2  + u_B|b_0|^2\bigg[|b_0|^2-1 \bigg]\bigg\},
\end{equation}
where the quartic contribution has been neglected. The temperature dependence of all constants can also be neglected and the expression above simplified to get
\begin{equation}
S_{eff}(\beta)\approx \beta \bigg\{\frac{|\Delta|^2}{|u_F|}-a|\Delta+g_rb^{ }_0|^2 + \left[2\nu-\mu_B\right]|b_0|^2  + u_B|b_0|^2\bigg[|b_0|^2-1 \bigg]\bigg\},
\end{equation}
The Helmholtz free energy at $T=0$ is calculated from 
\begin{equation}
 \Omega = \lim_{\beta\rightarrow\infty} -\frac{1}{\beta}\ln{Z(\beta)},
\end{equation}
where
\begin{equation}
\label{chempot_partfn}
Z(\beta) = \int D[b^*, b] D[\Delta^*, \Delta] e^{-S_{eff}(\beta)}.
\end{equation}
Simplifying equation~\ref{chempot_partfn} by taking the mean field (ignoring fluctuations) yields
\begin{equation}
\label{chempot_partfn_mf}
Z(\beta) \approx e^{-S_{eff}(\beta)}.
\end{equation}
Thus, the Helmholtz free energy $\Omega$ at $T=0$ is given by
\begin{equation}
\Omega \approx \frac{|\Delta|^2}{|u_F|}-a|\Delta+g_rb^{ }_0|^2 +\left[2\nu-\mu_B\right]|b_0|^2  + u_B|b_0|^2\bigg[|b_0|^2-1 \bigg].
\end{equation}
Imposing number conservation by using $\mathcal{N} = - {\partial\Omega}/{\partial\mu_F}$, 
\begin{equation}
\label{eq:numfb}
\mathcal{N} = \frac{\partial a}{\partial\mu} |\Delta + g_rb_0|^2 + 2 |b_0|^2.
\end{equation}
where $\mu=\mu_F=-\epsilon_F$, and the dependence of $a$ on $\mu$ is obtained from equations~\ref{Dl0_equation}. Writing the equation above  in terms of dimensionless variables, 
\begin{equation}
\label{chempot}
 2 \sigma^2|\Psi_2|^2 + \xi^2 |\Psi_1|^2 = 1, 
\end{equation}
where 
\begin{equation}
\label{xisq}
\xi^2 \equiv \mu^2\frac{\partial a}{\partial \mu}.
\end{equation}
Equation~\ref{chempot} is the constraint that fixes the number of particles $\mathcal{N}$ via the chemical potential $\mu_F$. Furthermore, in the stationary case, equations~\ref{dynamical_system} reduce to
\begin{eqnarray}
\label{fpcomplex}
\gamma \left(  \bar{\Psi}_1- \bar{\Psi}_2 \right) - \alpha  \bar{\Psi}_1 +\beta | \bar{\Psi}_1|^2  \bar{\Psi}_1 &=& 0, \nonumber \\
2\lambda_\nu  \bar{\Psi}_2 +2 \chi| \bar{\Psi}_2|^2 \bar{\Psi}_2 - \kappa\gamma \left( \bar{\Psi}_1- \bar{\Psi}_2 \right) &=& 0,
\end{eqnarray}
where $ \bar{\Psi}_1, \bar{\Psi}_2$ are the stationary solutions of the dynamical system. The trivial stationary solutions of the dynamics in the $\Psi_1,\Psi_2$ phase space, $ \bar{\Psi}_1, \bar{\Psi}_2=0$, are realized at temperatures above the critical temperature $T_c$~\cite{huang:bcsbecgp}. The nontrivial ones are the locus of points satisfying the equations above plus the chemical potential equation~\ref{chempot}. Thus, the 3 unknowns $ \bar{\Psi}_1, \bar{\Psi}_2,\epsilon_F$ are solved from the 3 simultaneous equations
\begin{eqnarray}
\label{fixedpoints}
\gamma\left(\epsilon_F\right) \left(  \bar{\Psi}_1- \bar{\Psi}_2 \right) - \alpha\left(\epsilon_F\right)  \bar{\Psi}_1 +\beta\left(\epsilon_F\right) | \bar{\Psi}_1|^2  \bar{\Psi}_1 &=& 0, \nonumber \\
2\lambda_\nu\left(\epsilon_F\right)  \bar{\Psi}_2 +2 \chi\left(\epsilon_F\right)| \bar{\Psi}_2|^2 \bar{\Psi}_2 - \kappa\left(\epsilon_F\right)\gamma\left(\epsilon_F\right) \left( \bar{\Psi}_1- \bar{\Psi}_2 \right) &=& 0, \nonumber \\
2 \sigma^2(\epsilon_F) | \bar{\Psi}_2|^2 + \xi^2(\epsilon_F) | \bar{\Psi}_1|^2 -1 &=& 0.
\end{eqnarray}
As explained in section~\ref{sec:dyneqns}, the BCS-dominant regime is not entirely accessible in this formalism. However, the predominantly BEC gapless regime can be obtained from equations~\ref{fixedpoints}. In that regime, $\Delta=0$ ie $ \bar{\Psi}_1= \bar{\Psi}_2$. In the case of noninteracting Bosons, ie $u_B=0$ ($\chi=0$), the second equation of~\ref{fixedpoints} necessitates that $\lambda_\nu=0$ ie $\epsilon_F=-\nu$, in agreement with mean field results~\cite{huang:becbcs2}. This is also consistent with the physics of the system, since the chemical potential is the energy required to remove one Fermion from the system. In the gapless and noninteracting BEC dominant case, the majority of the Fermions are dimerised with binding energy $-\nu$ , and a dimerised Fermion requires $\nu$ energy to dissociate from the dimer and free itself. For the more general case,
\begin{equation}
\label{bec_fp}
|\bar{\Psi}_{1,2}|^2 = -\frac{\lambda_\nu}{\chi}=\frac{\alpha}{\beta}.
\end{equation}
Clearly, $\lambda_\nu$ and so $\nu$ has to be negative for this to be true ($\chi$ is always positive). Also
\begin{equation}
\label{bec_fp2}
 \frac{\alpha}{\beta}+\frac{\lambda_\nu}{\chi}=0.
\end{equation}
Since $k_R \gg 1$, so is $\alpha$. Therefore, for the equation above to hold, $\lambda_\nu$ and therefore $\nu$ must be large and negative for a gapless BEC dominant state to exist. This criterion is in agreement with the physics of the system. For large negative $\nu$, the binding energy and therefore the molecular affinity of the Boson dimers will be large and negative, facilitating the dimerisation of the majority of the Fermions~\cite{bcsbec:ohashi}.
 In order to calculate the chemical potential in this regime, the relation $\alpha=\beta| \bar{\Psi}_2|^2$ is substituted in equation~\ref{bec_fp2}. Then,  equations~\ref{greek_consts_chempot} are applied, yielding 
\begin{equation}
 -\frac{\nu+\epsilon_F-\frac{1}{2}u_B}{u_B} = \frac{g_r |b_0|^2}{\epsilon_F}.
\end{equation}
Solving this quadratic equation in $\epsilon_F$ and rejecting the unphysical root, 
\begin{equation}
\label{chempot_purebec}
\epsilon_F = -\frac{1}{2}\left(\nu-\frac{1}{2}u_B\right)\bigg\{1+ \bigg[1-\frac{4g_ru_B|b_0|^2}{\left(\nu-\frac{1}{2}u_B\right)^2} \bigg]^{1/2}\bigg\} \approx - \left(\nu-\frac{1}{2}u_B\right) \bigg[1-\frac{g_ru_B|b_0|^2}{\left(\nu-\frac{1}{2}u_B\right)^2} \bigg].
\end{equation}
In the case of noninteracting Bosons, equation~\ref{chempot_purebec} gives $\epsilon_F=-\nu$ as above. Furthermore, in this regime $|b_0|^2=\mathcal{N}/2$ (no BCS state, all Fermions are dimerised), or $|\bar{\Psi}_{1,2}|^2 = 1/2\sigma^2(\epsilon_F)$. This simplifies equation~\ref{chempot} to $\xi = 0$ or  $\partial a/\partial\mu = 0$. Figure~\ref{fig:xi} shows plots of $\partial a/\partial\mu $ as functions of $\nu$ (where $\nu= \mu_F$) for $m=1$, $\mathcal{V}=1$ and several large values of $k_R$. Note that  $\partial a/\partial\mu $ and therefore $\xi$ indeed vanishes in the limit $\nu \rightarrow -\infty$, which is where the BEC regime is expected.

Continuing with the case of noninteracting Bosons ($u_B=0$ ie $\chi=0$), equations~\ref{fixedpoints} reduce to 
\begin{eqnarray}
\label{fixedpoints_noint}
\gamma\left(\epsilon_F\right) \left(  \bar{\Psi}_1- \bar{\Psi}_2 \right) - \alpha\left(\epsilon_F\right)  \bar{\Psi}_1 +\beta\left(\epsilon_F\right) | \bar{\Psi}_1|^2  \bar{\Psi}_1 &=& 0, \nonumber \\
\lambda_\nu\left(\epsilon_F\right)  \bar{\Psi}_2 - \kappa\left(\epsilon_F\right)\gamma\left(\epsilon_F\right) \left( \bar{\Psi}_1- \bar{\Psi}_2 \right) &=& 0, \nonumber \\
2 \sigma^2(\epsilon_F) | \bar{\Psi}_2|^2 + \xi^2(\epsilon_F) | \bar{\Psi}_1|^2 -1 &=& 0.
\end{eqnarray}
with $\lambda_\nu(\epsilon_F)=\frac{(2m\epsilon_F)^{3/2}}{32\pi}[1+\frac{\nu}{\epsilon_F}]$. Solving the first two equations of~\ref{fixedpoints_noint} yields
\begin{eqnarray}
\label{fixedpoints_noint_final}
 \bar{\Psi}_1 &=& \bigg[\frac{1}{\beta} \left(\alpha - \gamma\eta_\nu \right)\bigg]^{1/2} e^{i\theta} \nonumber \\
 \bar{\Psi}_2 &=& \left( 1-\eta_\nu\right)  \bar{\Psi}_1,
\end{eqnarray}
where $\eta_\nu=\frac{\lambda_\nu/\kappa\gamma}{1+\left(\lambda_\nu/\kappa\gamma\right)}$.  {The problem of renormalizability of this dual-channel theory mentioned in the previous section manifests here. The inability to renormalize happens because of the quantity $\alpha-\gamma\eta_\nu$ in the equations above. In the single channel case, when $g_r, \kappa\rightarrow0$ and $\eta_\nu\rightarrow 1$, this expression converges as $k_r\rightarrow \infty$, as can be seen using equations}~\ref{greek_consts_chempot},  {yielding}
\begin{equation}
\lim_{k_R\rightarrow\infty}\left[\lim_{g_r\rightarrow0}\left(\alpha-\eta\gamma\right)\right] = -\left[\frac{\epsilon_F}{|u^0_F|}+\frac{1}{8\pi}(2m\epsilon_F)^{3/2}\right].
\end{equation}
 {However, finite nonzero $g_r$ leads to $\eta$ vanishing as $k_R\rightarrow\infty$, yielding $\lim_{k_R\rightarrow\infty}\left(\alpha-\eta\gamma\right) \sim \alpha$ which diverges for arbitrarily large $k_R$} (see equations~\ref{greek_consts_chempot})\footnote{ {It should be noted, however, that this divergence goes away at $T\neq 0$ due to the use of thermal Green's functions (instead of the ones from equation}~\ref{greens:function})  {in the evaluation of the constants in equations}~\ref{abc} ~\cite{machida:dynamics}~\cite{huang:bcsbecgp}.}.

 {The difficulty described above can be worked around by noting that the allowed momenta need to have a finite cutoff $k_R$, which is governed by the physics of the system. For instance, in the case of solid state BCS systems, the cutoff is given by the Debye frequency on account of the physical origins of the electron-electron attraction in the slowly relaxing lattice vibrations. Ordinarily, the choice of cutoff is decided by the range of the interatomic potential, with $k_R\sim 8\pi^2\nu/g^2_r$ if the bare repulsion between the Fermions is neglected~\cite{timmermans,pethick:bec}. However, for small confinements and small detuning, the maximum momentum allowed in these cold atom systems cannot be greater than $2\pi/R$, where $R$ is the size of the atoms. Experimentally, these atoms are very cold, and never move so fast as to cause inelastic scattering beyond the Feshbach resonance, since such excitations will dynamically alter the internal degrees of freedom of the atoms themselves. Thus, this choice of $k_R$ keeps a reasonable upper bound in all of the integrations. For Fermionic atoms like $Li$, $R$ is ${\mathcal O}(10^{-10}m)$, which needs to be rescaled with respect to the unit length viz. the confinement dimensions. The confinement, however, is highly tunable in cold atom systems, with a fairly wide range of permitted values. Thus, using a wide variety of $k_R$'s to get quantitatively correct results should be permissible, once the trap sizes are adjusted accordingly. In this report,  $k_R$ are chosen to be ${\mathcal O}(10)$ in units of inverse trap size. These should be attainable in small volume systems of ${\mathcal O}(10^2)$ atoms (obtained experimentally via laser culling by Chuu et al}~\cite{raizen:culling})  {confined to length scales of ${\mathcal O}(10^{-9}m)$  (obtained experimentally by confining ultracold gases using atom chips}~\cite{doublewell:chip}).

\begin{figure}[h!bt]
\hspace*{-0.3in}
\ \epsfig{file=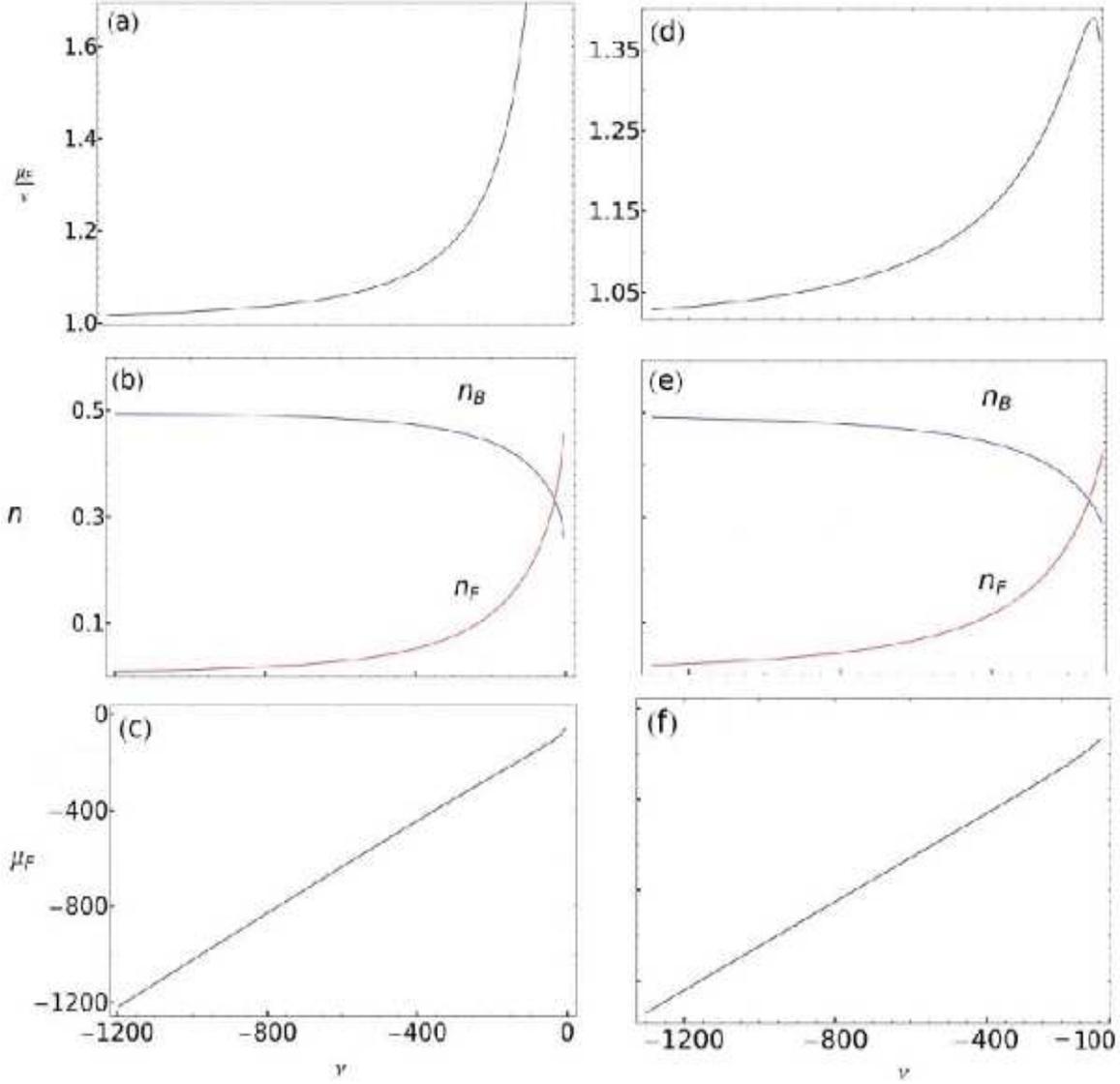,height=6in,width=6in}
\caption{(Color online) Plots of the  Fermion/Boson number and  chemical potential $\mu_F$  as a function of the Feshbach detuning $\nu$ in the deep-BEC and shallow-BEC regimes. Here, $\hbar = m =\mathcal{V} = 1$. The renormalization cutoff $k_R$ is set to $18$. The curves are evaluated for two values of the atom molecule coupling $g_r$. Figures (a), (b) and (c) contain plots for the ratio $\mu_F/\nu$, the condensate fractions and $\mu_F$ respectively for $g_r=25$, $u_F = -0.3$, $u_B = 0$  and $\mathcal{N} = 100$. Figures (d), (e) and (f) contain plots for the ratio $\mu_F/\nu$, the condensate fractions and $\mu_F$ respectively for $g_r=40$, $u_F = -0.3$, $u_B = 0$  and $\mathcal{N} = 100$. In figures (b) and (e), the Fermion condensate fraction $n_F$ is shown in red, and the Boson molecular condensate fraction $n_B$  is shown in blue. Note that the pure molecular BEC is achieved when $n_B = 0.5$, half the total number $\mathcal{N}$. Also, note the onset of the shallow BEC regime at $\nu \approx -50$ for (b) and at $\nu \approx -150$ for (e). Also, note from figures (a) and (d) that $\mu_F$ approaches $\nu$ for sufficiently large values of $-\nu$, indicating the onset of a pure molecular BEC in accordance with the results in Equation~\ref{chempot_purebec}. Finally, note that figs (c) and (f) show that $\mu_F$ is approximately equal to $\nu$ for large $|\nu|$, but deviates from this at smaller $|\nu|$, as can be seen more clearly in figs (a) and (d).}
\label{fig:chempot}
\end{figure}

Equations~\ref{fixedpoints_noint_final} are nothing more than the gap equation for the BCS-BEC system in position space. In all these equations, $\alpha,\beta,\gamma,\kappa,\lambda_\nu,\eta_\nu$ are functions of $\epsilon_F$. Note from the above that if $\kappa\rightarrow0$ ie $g_r \rightarrow0$, then $\eta_\nu\rightarrow 1$, and  $ \bar{\Psi}_2\rightarrow0$, which will give  a pure BCS until $\epsilon_F$ goes to a regime where $\alpha$ can no longer be evaluated without running into singularities (as per section~\ref{sec:dyneqns}), causing this formalism to break down. Plugging the values of $\bar{\Psi}_{1,2}$ above to the final equation of~\ref{fixedpoints_noint} yields
\begin{equation}
\label{chempot_final_noint}
1=\frac{1}{\beta(\epsilon_F)}\bigg[\alpha(\epsilon_F) - \gamma(\epsilon_F)\eta_\nu(\epsilon_F) \bigg] \bigg\{2\sigma^2(\epsilon_F)\big[1-\eta_\nu(\epsilon_F)\big]^2 +\xi^2(\epsilon_F)\bigg\}.
\end{equation}
\pagebreak
Equation~\ref{chempot_final_noint} is a transcendental equation and needs to be solved numerically for $\mu_F$ ($-\epsilon_F$). 
Figure~\ref{fig:chempot} contains the numerical results of evaluating the condensate fractions and chemical potentials. These have been obtained by setting $\mathcal{N}$ to $100$ and using representative values of the interaction parameters. The chemical potential $\mu_F$ was obtained by numerically solving equation~\ref{chempot_final_noint} using Newton-Raphson methods. The default working precision was kept to the eighth place of decimal. Figures~\ref{fig:chempot} (a), (b) and (c) contain plots for $g_r = 25$, and figures~\ref{fig:chempot} (d), (e) and (f) contain plots for $g_r = 40$. Figures~\ref{fig:chempot}(a) and (d) are plots for $\mu_F/\nu$ as a function of $\nu$. Note from these figures that, for sufficiently large $-\nu$, $\mu_F \approx \nu$ in accordance with the analytical results in the BEC regime. Figures~\ref{fig:chempot}(b) and (e) are plots for the condensate fractions $n_F$ (Fermions in BCS) and $n_B$ (molecular composite Bosons in BEC). The fractions are computed after solving equation~\ref{chempot_final_noint}, and substituting the corresponding values of $|\bar{\Psi}_{1,2}|^2$ (obtained from
equations~\ref{fixedpoints_noint_final}) into the relations (see equation~\ref{chempot})
\begin{eqnarray}
n_F &=& \xi^2 | \bar{\Psi}_1|^2, \nonumber \\
n_B &=& \sigma^2| \bar{\Psi}_2|^2.
\end{eqnarray}
The rise in the BCS superfluid density can be clearly seen in figure~\ref{fig:chempot}(b) around $\nu \approx -50$, and in figure~\ref{fig:chempot}(e) of $\nu \approx -150$. These results agree qualitatively with more sophisticated theories of BCS-BEC systems~\cite{leggett:book}, considering that numbers are highly sensitive to the choice of the renormalization cutoff $k_R$.
In the case of interacting bosons, equations~\ref{fixedpoints} are simplified by operating in a regime where the chemical potential is weak enough so that $\beta$ can be ignored. This yields
\begin{eqnarray}
\label{fixedpoints_nobeta}
\gamma\left(\epsilon_F\right) \left(  \bar{\Psi}_1- \bar{\Psi}_2 \right) - \alpha\left(\epsilon_F\right)  \bar{\Psi}_1 &=& 0, \nonumber \\
2\lambda_\nu\left(\epsilon_F\right)  \bar{\Psi}_2 +2 \chi\left(\epsilon_F\right)| \bar{\Psi}_2|^2 \bar{\Psi}_2 - \kappa\left(\epsilon_F\right)\gamma\left(\epsilon_F\right) \left( \bar{\Psi}_1- \bar{\Psi}_2 \right) &=& 0, \nonumber \\
2 \sigma^2(\epsilon_F) | \bar{\Psi}_2|^2 + \xi^2(\epsilon_F) | \bar{\Psi}_1|^2 -1 &=& 0.
\end{eqnarray}
Solving the first two equations of~\ref{fixedpoints_nobeta} yields the gap equations in this regime. The solutions are
\begin{eqnarray}
\label{fixedpoints_nobeta_final}
\bar{\Psi}_1 &=& \frac{\alpha}{\gamma-\alpha} \bar{\Psi}_2, \nonumber \\
\bar{\Psi}_2 &=& \bigg \{-\frac{\lambda_\nu}{\chi} \bigg[1+\frac{\kappa\gamma}{2\lambda_\nu\left(\gamma-\alpha \right)} \bigg]  \bigg \}^{1/2} e^{i\theta},
\end{eqnarray}
where a necessary condition is that $\lambda_\nu$ be negative. Substituting these results into the final equation of~\ref{fixedpoints_nobeta} yields the transcendental equation for $\epsilon_F$ in this case,
\begin{equation}
\label{chempot_final_nobeta}
1+\frac{2\sigma^2\left(\epsilon_F\right)\lambda_\nu\left(\epsilon_F\right)}{\chi\left(\epsilon_F\right)} \bigg\{1+\frac{\xi^2\left(\epsilon_F\right)\gamma^2\left(\epsilon_F\right)}{2\sigma^2\left(\epsilon_F\right)\left[\gamma\left(\epsilon_F\right)-\alpha\left(\epsilon_F\right) \right]^2} \bigg\} \bigg\{1+\frac{\kappa\left(\epsilon_F\right)\gamma\left(\epsilon_F\right)}{2\lambda_\nu\left(\epsilon_F\right)\left[\gamma\left(\epsilon_F\right)-\alpha\left(\epsilon_F\right) \right]} \bigg\} = 0.
\end{equation}
\pagebreak
\pagebreak
\begin{figure}[h!bt]
\ \epsfig{file=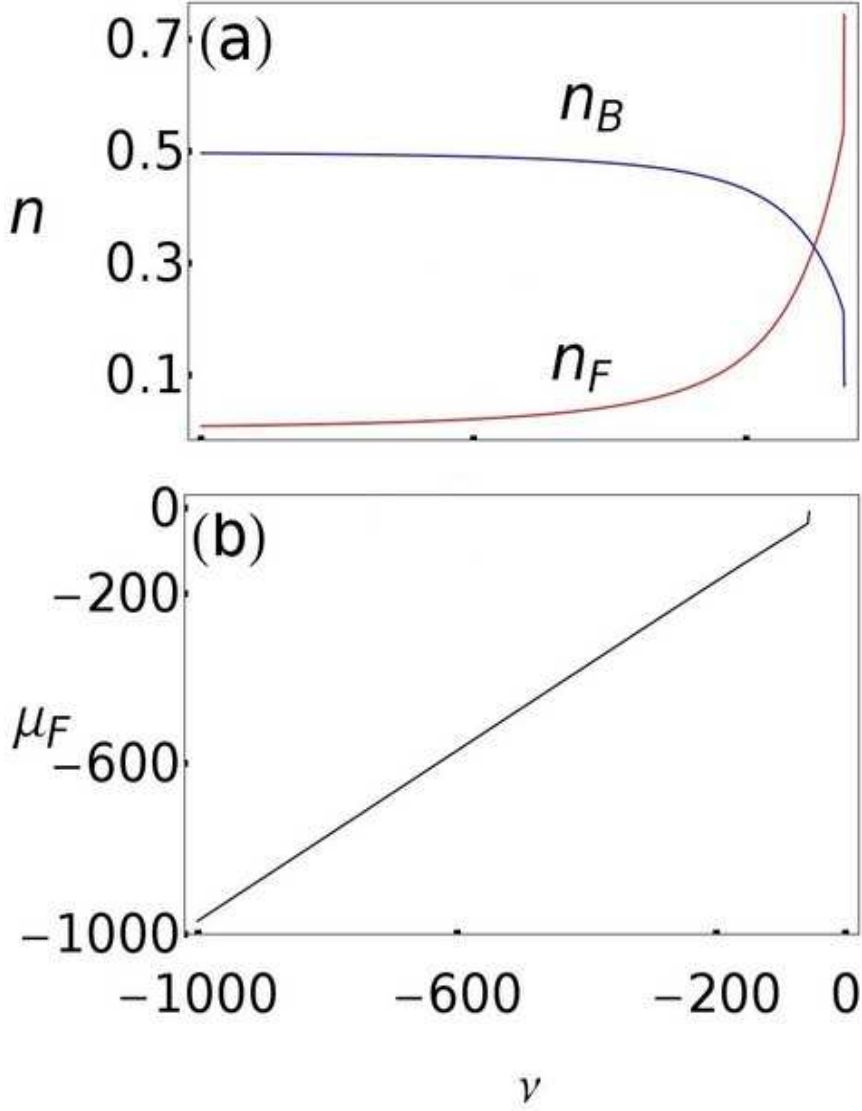,height=6in,width=4.5in}
\caption{(Color online) Plots of the  Fermion/Boson number and  chemical potential $\mu_F$  as a function of the Feshbach detuning $\nu$ in the deep-BEC and shallow-BEC regimes. Here, $\hbar = m =\mathcal{V} = 1$, $u_F= -0.3$, $u_B = 0.6$ and $\mathcal{N}=100$. The renormalization cutoff $k_R$ is set to $14.5$. The curves are evaluated for $g_r=25.0$. Figure (a) contains plots for the condensate fractions as a function of the Feshbach detuning $\nu$. The Fermion condensate fraction $n_F$ is shown in red, and the Boson molecular condensate fraction $n_B$  is shown in blue. Note that the pure molecular BEC is achieved when $n_B = 0.5$, half the total number $\mathcal{N}$. Also, note the onset of the shallow-BEC regime at $\nu \approx -100$. Figure (b) contains plots of the chemical potential $\mu_F$ as a function of $\nu$. Note here that $\mu_F$ approaches $\nu$ for sufficiently large values of $-\nu$, indicating the onset of a pure molecular BEC in accordance with the results in Equation~\ref{chempot_purebec}.}
\label{fig:chempot2}
\end{figure}
\pagebreak
\begin{figure}[h!bt]
\ \epsfig{file=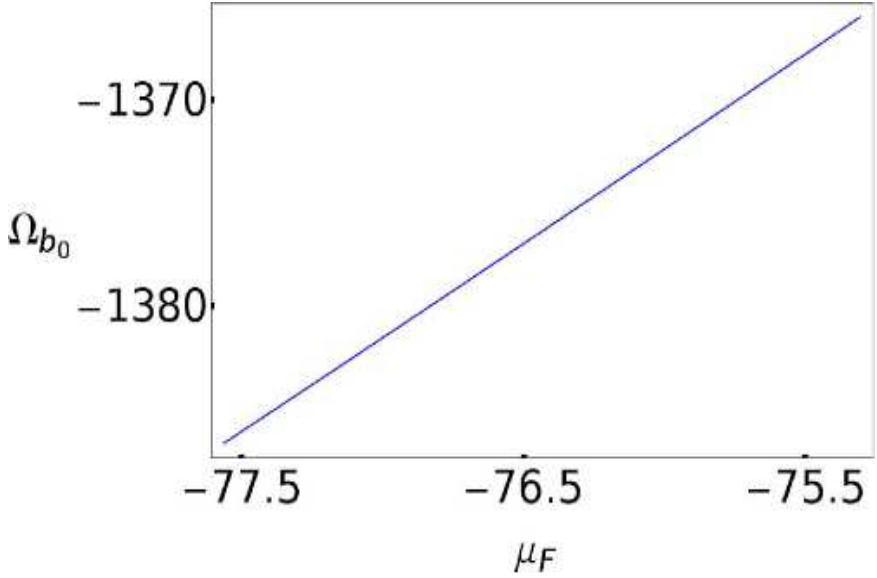,height=3.0in,width=4.5in}
\caption{(Color online) Plots of $\Omega_{b_0}$ as a function of chemical potential $\mu_F$ for $g_r=40$, $u_F=-0.3$, $u_B=0$, and $\mathcal{N}=100$. Compare order of magnitudes with that of $\Omega_m$ in Table~\ref{tab:colrev}.}
\label{fig:omega}
\end{figure}
Equation~\ref{chempot_final_nobeta} is solved numerically using the same algorithms and tolerances as equation~\ref{chempot_final_noint} for representative values of the parameters. The results are shown in figures~\ref{fig:chempot2}. Figure~\ref{fig:chempot2} (a) contains plots of the condensate fractions $n_F$ (red) and $n_B$ (blue) for $\mathcal{N}=100$, $g_r = 25$, $|u_F| = 0.3$, and $u_B = 0.6$. Here, $k_R$ is chosen to be $14.5$. Note the shallow BEC point at $\nu \approx -100$ in Figure~\ref{fig:chempot2} (a). In this case, the fermion population appears to increase much faster than in the case of noninteracting Bosons as $\nu$ is varied adiabatically past this regime. The system should populate to a full BCS state shortly after unitarity.
\section{Quenched Time Evolution}
\label{sec:quench}
Equations~\ref{dynamical_system} provide a complete description of mean field dynamics of the Timmermans' Hamiltonian in equation~\ref{timmermans}. The $\Psi_i$s will not, in general, represent densities when the system departs from equilibrium. The definition of $\Psi_2$ in equation~\ref{vartrans} indicates that  while $|\Psi_2|^2(t)$ will be proportional to the Boson density, $|\Psi_1|^2(t)$ will no longer represent the Fermion density out of equilibrium (which will only depend on $1-2\sigma^2|\Psi_2|^2$), since equation~\ref{eq:numfb} breaks down away from equilibrium. The general quasiharmonic solutions of $\Psi_{1,2}$ have been discussed by Machida and Koyama~\cite{machida:dynamics}, and the global existence of weak solutions have been established recently by Chen \textit{et al}~\cite{chen:tdglsoln}. This paper focuses on the dynamics in the shallow BEC regime close to the BCS-BEC crossover as the system is quenched from the deep-BEC regime. 

The continued use of the single mode approximation in this case can be justified by evaluating the maximum characteristic time scale $t_q$ for the quench below which sizable excitations of the closed channel Bosons and free Fermions will occur. In order to do so, the Timmermans' Hamiltonian in equation~\ref{timmermans} is written in momentum space with the closed channel excitations included, yielding
\begin{multline}
\label{eq:timmermans:kspace}
H_{tm} = \sum_{{\bf k},\sigma} \left(\epsilon_{\bf k} - \mu_F \right)a^\dagger_{{\bf k}\sigma}a^{ }_{{\bf k}\sigma} - |u_F|\sum_{{\bf k}{\bf k'}}a^\dagger_{{\bf k}\uparrow}a^\dagger_{-{\bf k}\downarrow}a^{ }_{{\bf k'}\downarrow}a^{ }_{{\bf k'}\uparrow}\\
+\sum_{\bf q} \left(E^0_{\bf q}+2\nu-\mu_B \right)b^\dagger_{\bf q}b^{ }_{\bf q} + u_B\sum_{{\bf q_1}{\bf q'_1} {\bf q_2} {\bf q'_2}}b^\dagger_{\bf q'_1}b^\dagger_{\bf q'_2} b_{\bf q_2} b_{\bf q_1}\delta_{{\bf q_1}+{\bf q_2},{\bf q'_1}+{\bf q'_2}}\\
+g_r\sum_{{\bf k}{\bf q}}\left(b^\dagger_{\bf q}a_{{\bf p+\frac{q}{2}}\uparrow}a_{{\bf -p+\frac{q}{2}}\downarrow} + h.c. \right).
\end{multline}
Here, the first line represents the Fermions in momentum space, with the creation (annihilation) operators $a^\dagger_{{\bf p}\sigma}$ ($a^{ }_{{\bf p}\sigma}$) obtained by Fourier transforms of $\phi^\dagger_\sigma$ ($\phi^{ }_\sigma$) from equation~\ref{timmermans}, and the free Fermion energies are represented by $\epsilon_{\bf k}$. The second line represents the Hamiltonian for the closed channel quasimolecular bosons $b^\dagger_{\bf q}$ ($b^{ }_{\bf q}$) with energies $E^0_{\bf q}+2\nu-\mu_B$. Finally, the last line represents the atom-molecule coupling discussed in section~\ref{sec:dyneqns}. The Fermi and Bose condensates exchange momentum ${\bf q}$ through this coupling. The Hamiltonian above can be written in terms of the generators of the $SU(2)\otimes SU(1,1)$ Lie algebra~\cite{huang:becbcs2}. Assuming that most of the Bosons are in the ground state given by ${\bf q}=0$, treating $b_0$ as a c-number, retaining only terms up to second order in $b^\dagger_{\bf q}$ ($b^{ }_{\bf q}$) and neglecting fluctuations about the mean field for all generators, the Hamiltonian can be diagonalized in a manner similar to that done for a pure Boson system in Bogoliubov theory~\cite{huang:becbcs2}, with the eigenstates given by the generalized $SU(2)\otimes SU(1,1)$ coherent state. If the system is in a pure quantum state, the energy eigenvalues are given by~\cite{huang:becbcs2}
\begin{equation}
E_{sn} = \left(2\nu-\mu_B \right)|b^2_0| + \sum_{\bf k} \left(2sE_{\bf k} +\epsilon_{\bf k}-\mu_F \right) +\sum_{\bf k\neq 0} \left[ \left( n+\frac{1}{2}\right)E^b_{\bf k} -\frac{1}{2}\left(E^b_{\bf k} + 2\nu-\mu_B+2u_B|b_0|^2\right)\right].  
\end{equation}
Here, $E_{\bf k} = \sqrt{(\epsilon_{\bf k}-\mu_F)^2+|\Delta+g_rb_0|^2}$ are the energies of the Bogoliubov quasiparticles from conventional BCS theory, $E^b_{\bf k} = \sqrt{E^0_{\bf k}(E^0_{\bf k} + 2u_B |b_0|^2)}$ are the energies of the quasiparticle excitations of  the pure Boson system from Bogoliubov theory~\cite{huang:becbcs2}, and $s$/$n$ are indices that correspond to the ladder operators of the $SU(1,1)$/$SU(2)$ Lie algebras respectively, with $s=\pm\frac{1}{2}$ and $n\in\mathbb{N}^0$. If the energies are referenced from the ground state $E_{-\frac{1}{2}0}$, then the excitation energies are given by
\begin{eqnarray}
\label{eq:excitations}
 \delta E_{\frac{1}{2}n}&=& 2\sqrt{\mu^2_F + |\Delta + g_rb_0|^2} + \sum_{{\bf p}\neq 0}
\left(2E_{\bf p}+nE^b_{\bf p} \right),  \nonumber \\
\delta E_{-\frac{1}{2}n} &=& \sum_{{\bf p}\neq0} nE^b_{\bf p}, 
\end{eqnarray}
where $\epsilon_{{\bf k}=0}$ is taken to be $0$. Evaluating equation~\ref{eq:excitations} in $3$D with $\epsilon_{\bf k} = k^2/2m$, $E^0_{\bf p}=p^2/4m$ and a momentum cutoff at $p=p_c\equiv\frac{2\pi}{R}$ to counter ultraviolet divergences shows that the sums in these equations go as positive powers of the particle number density (noting that $\mathcal{V}=1$ here) $\rho\sim u_B |b_0|^2\sim\frac{1}{R^3}$, in particular,  $\delta E_{-\frac{1}{2}n}\sim \rho^{5/3}$. Thus, a lower bound for the excitation energies is given by 
\begin{equation}
E_c = 2\sqrt{\mu^2_F + |\Delta + g_rb_0|^2}.
\end{equation}
$E_c\sim \rho^{1/2}$ (see equation~\ref{eq:numfb}), which is fairly macroscopic due to the presence of the effective gap $\Delta + g_rb_0$. The corresponding relaxation time $t_c\equiv 1/E_c$ defines the time scale below which excitations of energy greater than $E_c$ are imparted, and  excitations away from the ground state (such as closed channel bosons and unpaired Fermions dissociated from them) are possible. In all subsequent analyses, it is assumed that the quench time $t_q \leq t_c$ and both are neglected in comparison to the time scales of the post quench dynamics. This effectively makes the quench an instantaneous impulse, and an analysis of the dynamical system based only on a Boson field and Fermion pair field completely captures the non-equilibrium state.

If there is no BCS state in this system for finite values of $\epsilon_F$, then $\Psi_1 = \Psi_2 \equiv \Psi$, and the dynamics is equivalent to one with no atom-molecule coupling $g_r$. Thus, equations~\ref{dynamical_system} simplify to
\begin{eqnarray}
\label{dynamical_system_nogr}
\frac{\partial\Psi}{\partial t} - i\alpha \Psi +i\beta |\Psi|^2 \Psi &=& 0, \nonumber \\
\frac{\partial\Psi}{\partial t} +2 i \lambda_\nu \Psi +2 i\chi|\Psi|^2\Psi &=& 0.
\end{eqnarray}
Substituting quasiharmonic trial solutions $\Psi = \tilde{\Psi} e^{i\Omega t}$ yields
\begin{eqnarray}
\label{dynamical_system_nogr_ft}
\Omega \tilde{\Psi} - \alpha \tilde{\Psi} +\beta |\tilde{\Psi}|^2 \tilde{\Psi} &=& 0, \nonumber \\
\Omega\tilde{\Psi} +2  \lambda_\nu \tilde{\Psi} +2 \chi|\tilde{\Psi}|^2\tilde{\Psi} &=& 0.
\end{eqnarray}
Both of these equations need to hold for two unknowns $\tilde{\Psi}$, $\Omega$. Thus, in the absence of Boson interactions ($\chi = 0$), $\Omega=- 2 \lambda_\nu$, and the phase of the Boson condensate fraction oscillates without any population transfer out of the BEC in all regimes.

The shallow-BEC regime is one where the condensate fractions $n_F$ and $n_B$ are comparable (see figure~\ref{fig:chempot}). If no \emph{a-priori} assumptions are made about any of the other constants, equations~\ref{greek_consts_chempot} indicate that $\beta$ can be ignored. In the case of noninteracting Bosons, a quench from a pure BEC, where $\psi_{1,2}(0) = \left(1/\sqrt{2} \right)g_r/\epsilon_F$, to this regime (where the dynamics is linear) produces \begin{eqnarray}
\label{dynamical_system_nobeta}
\frac{\partial\Psi_1}{\partial t}+ i\gamma \left( \Psi_1-\Psi_2 \right) - i\alpha \Psi_1 &=& 0, \nonumber \\
\frac{\partial\Psi_2}{\partial t} +2i\lambda_\nu\Psi_2 - i\kappa\gamma \left(\Psi_1-\Psi_2 \right) &=& 0.
\end{eqnarray}
Substituting trial solutions $\Psi_{1,2}(t) = \tilde{\Psi}_{1,2} \times e^{i\Omega t}$, and solving the  resulting characteristic equation for $\Omega$,  
\begin{equation}
\label{omegapm}
\Omega_\pm\left(\epsilon_F \right) = -\frac{1}{2} {\mathrm A}\left(\epsilon_F\right) \bigg[ 1 \pm \sqrt{1-4\frac{{\mathrm B}\left(\epsilon_F\right)}{{\mathrm A}^2\left(\epsilon_F\right)}}\bigg]. 
\end{equation}
where
\begin{eqnarray}
\label{AB}
{\mathrm A}\left(\epsilon_F\right) &\equiv& \gamma -\alpha +2\lambda -\kappa\gamma , \nonumber \\
{\mathrm B}\left(\epsilon_F\right) &\equiv& \left(\gamma-\alpha \right)\left(2\lambda+\kappa\gamma \right)-\kappa\gamma^2.                                                                      
\end{eqnarray}
Here, as always the constants in Greek letters are functions of $\epsilon_F$. It is noted that, for sufficiently large $k_R$, $\mathrm{B}(\epsilon_F)$ is negative, and so $\Omega_\pm$ are both real (no damping or exponential growth). The BEC condensate fraction,  proportional to $|\Psi_2|^2(t)$,  oscillates with frequency $\Omega_{b_0} = |\Omega_+ - \Omega_-|$. Substituting the values above, 
\begin{equation}
\Omega_{b_0} = {\mathrm A}\left(\epsilon_F\right)\sqrt{1-4\frac{{\mathrm B}\left(\epsilon_F\right)}{{\mathrm A}^2\left(\epsilon_F\right)}}.
\end{equation}
For sufficiently large cutoff frequency $k_R$, $\Omega_{b_0} \sim k_R$ and thus it is believed to represent Rabi oscillations of solitons that are similar to the ones reported in $2004$ by Andreev, Gurarie, and Radzihovsky~\cite{andreev:noneqmbcsbec}, as well as others since $2004$~\cite{barankov:bcsbecbloch,yuzbashyan2,miyakawa} for a quantum quenched noninteracting Fermi gas with a Feshbach resonance. Numerical results for $\Omega_{b_0}$ are plotted for $g_r=40$ in figure~\ref{fig:omega}. 

A linear stability analysis of small displacements away from equilibrium should provide some qualitative insight into the long term dynamics of the system even for large displacements from equilibrium, such as impulse quenches. Absent the interference term $\Psi_1-\Psi_2$, eqns~\ref{dynamical_system} resemble the normal form of the Hopf bifurcation~\cite{strogatz:book}. However, this does not actually occur due to the complex nature of the coefficients. Writing the dynamical variables in polar coordinates as $\Psi_j = a_j e^{i\phi_j}$ and dropping $\Psi_1-\Psi_2$ simplifies the dynamics to yield $\dot{a}_j=0$ and $\phi_j\propto t$, trivial trajectories that do not admit to bifurcations of any kind. 

Investigating the more general cases, however, do yield interesting results. Substituting polar coordinates into the dynamics in eqns~\ref{dynamical_system}, and splitting up the real and imaginary parts, results in a nonlinear dynamical system evolving in a $4$-dimensional phase space spanned by $a_{1,2}$ and $\phi_{1,2}$ viz.
\begin{eqnarray}
\label{dynamics:radial}
\dot{a}_1 &=& -\gamma a_2 \sin{\left(\phi_2-\phi_1\right)},\nonumber \\
\dot{a}_2 &=& \kappa \gamma a_2 \sin{\left(\phi_2-\phi_1\right)},\nonumber \\
\dot{\phi}_1 &=& \alpha - \beta a^2_1 - \gamma\left[1-\frac{a_2}{a_1}\cos{\left(\phi_2-\phi_1\right)} \right],\nonumber \\
\dot{\phi}_2&=& \kappa\gamma\left[\frac{a_1}{a_2}-\cos{\left(\phi_2-\phi_1\right)} \right]-2\lambda_\nu.
\end{eqnarray}
Here, only the first of two cases discussed in the previous section are considered, where the boson repulsion term $\chi$ has been ignored. The radial fixed points of $\dot{a}_j=0$ can be realized for nonzero $a_j$ if $\phi_2(t) = \phi_1(t)+n\pi$.  Note that, in this limit, $\cos{\left(\phi_2-\phi_1\right)}=\pm 1$ and the rhs of the last two of eqns~\ref{dynamics:radial} must be equal to each other. This is satisfied by eqn~\ref{chempot_final_noint}, once the fixed point values of $a_j$ from eqns~\ref{fixedpoints_noint_final} are taken as $a_1=r_\nu$, $a_2=(1-\eta_\nu)r_\nu$, with $r_\nu \equiv\left[\frac{1}{\beta}\left(\alpha -\gamma\eta_\nu \right)\right]^{1/2}$. Defining $\omega_\nu\equiv\dot{\phi}_1=\dot{\phi}_2$, and applying these expressions to the fixed points above, the system evolves at the radial fixed points as
\begin{eqnarray}
 \Psi_1(t) &=& r_\nu e^{-i\omega_\nu t},\nonumber \\
 \Psi_2(t) &=& (1-\eta_\nu)r_\nu e^{-i\omega_\nu t},\nonumber \\
\omega_\nu &=& \bigg\{\begin{array}{l}
                         \lambda_\nu\\
			 \lambda_\nu - 2\kappa\gamma 
                       \end{array}.
\end{eqnarray}
Small displacements away from this orbit can be provided with complex quantities $\delta\Psi_j$ as follows. 
\begin{eqnarray}
 \Psi_1(t) &=& \left[r_\nu + \delta\Psi_1(t)\right]e^{-i\omega_\nu t},\nonumber \\
 \Psi_2(t) &=& \left[\left(1-\eta_\nu\right)r_\nu + \delta\Psi_1(t)\right]e^{-i\omega_\nu t}.
\end{eqnarray}
Substituting this into the dynamics in eqns~\ref{dynamical_system} yields
\begin{eqnarray}
 \delta\dot{\Psi}_1 &=& -i\left(\omega_\nu+\gamma-\alpha+2\beta r^2_\nu\right)\delta\Psi_1-i\beta r^2_\nu \delta\Psi^\ast_1+i\gamma\delta\Psi_2 + C_1,\nonumber \\
\delta\dot{\Psi}_2 &=& i\kappa\gamma\delta\Psi_1 -i\left(\omega_\nu+2\lambda+\kappa\gamma\right)\delta\Psi_2 + C_2,
\end{eqnarray}
where $C_j$ are constants, and terms $\mathcal{O}(|\delta\Psi_j|^2)$ and higher have been dropped. Separating the system into real and imaginary parts yields $\delta \dot{x}_j \sim \sum_k \mathcal{J}^{jk}_\nu\delta x_k$, where $\delta\Psi_{1,2}=\delta x_{1,3} + i \delta x_{2,4}$, and ${j,k}=1-4$. The Jacobian matrix elements that describe this dynamics are given by $\mathcal{J}^{jk}_\nu\equiv\partial\dot{x}_j/\partial x_k$. The matrix evaluates to
\begin{equation}
\label{jacobian:nobosonint}
\mathcal{J}_\nu \approx \begin{pmatrix}
 0 & -[\gamma -\alpha]-3 \beta r^2_\nu  & 0 & \kappa \gamma  \\
 [\gamma -\alpha ]+\beta r^2_\nu  & 0 & \kappa \gamma  & 0 \\
 0 & \gamma  & 0 & -\kappa\gamma \left[\frac{\eta_\nu+1}{\eta_\nu-1}\right] \\
 -\gamma  & 0 & -\kappa\gamma\left[\frac{\eta_\nu-3}{\eta_\nu-1}\right] & 0
\end{pmatrix},
\end{equation}
at the fixed points in eqn~\ref{fixedpoints_noint_final}. The eigenvalues $\Omega_\nu$ are given by the characteristic equation $|\mathcal{J}_\nu-\Omega_\nu I|=0$. In this case, the equation is biquadratic and yields $\Omega_\nu^4 + \mathcal{B}_\nu\Omega_\nu^2-\mathcal{C}_\nu =0$, where
\begin{eqnarray}
\label{roots:jacobian:nobosonint}
\Omega_\nu &=& \pm\sqrt{-\frac{\mathcal{B}_\nu}{2}}\left(1\pm\sqrt{\mathcal{D}_\nu}\right)^{1/2},\nonumber \\
 \mathcal{D}_\nu&\equiv& 1+\frac{4\mathcal{C}_\nu}{\mathcal{B}_\nu^2},\\
\label{nobosonint:coeff}
\mathcal{B}_\nu &\equiv& \beta r_\nu^2 \left[2\left(\alpha-\gamma\right)+3\beta r_\nu^2 \right]-\left(\alpha-\gamma\right)^2-
\frac{\left(\eta_\nu-3\right)\left(\eta_\nu+1\right)}{\left(\eta_\nu-1\right)^2}\times\kappa^2\gamma^2,\nonumber \\
\mathcal{C}_\nu &\equiv& \kappa^2\gamma^4+\frac{\left(\eta_\nu-3\right)\left(\eta_\nu+1\right)}{\left(\eta_\nu-1\right)^2}
\left[\left(\alpha-\gamma-2r^2_\nu\beta\right)^2-r^4_\nu\beta^2 \right]\kappa^2\gamma^2+2\kappa^2\gamma^3\left[\alpha-\gamma-\left(\frac{2\eta_\nu-1}{\eta_\nu-1}\right)r^2_\nu\beta\right].\nonumber \\.
\end{eqnarray}
The quantity $\mathcal{D}_\nu$ is the \textit{discriminant} of the quartic equation. The linear dynamics is given by~\cite{strogatz:book} $|\delta x (t) \rangle \sim \sum^{4}_{j=1} c_j |\Omega^j_\nu\rangle e^{\Omega^j_\nu t}$, where $c_j$ are constants, $|\Omega^j_\nu\rangle$ are the eigenvectors of $\mathcal{J}_\nu$  that correspond to eigenvalues $\Omega^j_\nu$, and $|\delta x (t) \rangle$ is the vector given by $\delta x_{1-4}$. The sign of the quartic discriminant $\mathcal{D}_\nu$ plays a key role in determining the nature of these trajectories. The regions of interest and the nature of the roots $\Omega_\nu$ are tabulated in table~\ref{tab:roots}. If $\Omega_\nu$ are pure imaginary, the trajectories correspond to orbital motion about the radial fixed points. The orbits constitute two oscillations of frequencies $\Omega^{\pm}_\nu=|\sqrt{-{\mathcal{B}_\nu}/{2}}(1\pm\sqrt{\mathcal{D}_\nu})^{1/2}|$. Their combination results in periodicity iff $\Omega^\pm_\nu$ have rational winding numbers $n^+/n^- = \Omega^+_\nu/\Omega^-_\nu$, where $n^\pm$ are integers with no common factors. In that case, the period of the resultant oscillations are $T_\nu = \pi (n^+/\Omega^+_\nu + n^-/\Omega^-_\nu)$. The changes in the number density of the BEC, denoted by $|\delta \Psi_2|^2=\delta x^2_3 + \delta x^2_4$, will have $4$ oscillations viz $2\Omega^\pm_\nu$ and $\Omega^+_\nu\pm\Omega^-_\nu$. The net period is much smaller than that of the $\Psi_j$s. When the $\Omega_\nu$ start to pick up real parts, then the trajectories corresponding to negative real parts are stable and decay in spirals to the orbits of the radial fixed points, thus creating a stable limit cycle. 
\begin{table}
 \begin{center}
\begin{tabular}{|l|l|l|l|}
\hline
		& $\mathcal{D}_\nu\leq0$    		             & $0<\mathcal{D}_\nu<1$		    &$\mathcal{D}_\nu											       		     \geq1$\\ 
\hline
		& $\Omega_\nu = \pm\sqrt{\frac{|\mathcal{B}_\nu|}{2}} 
		  \left(1\pm i\sqrt{|\mathcal{D}_\nu|}\right)^{1/2}$ & $\Omega_\nu=\pm\sqrt{\frac{|\mathcal{B}_\nu|}{2}} 
								\left(1\pm \sqrt{|\mathcal{D}_\nu|}\right)
												     ^{1/2}$ 
													    & $\Omega_\nu=\pm 														    \mbox{ }
													    \sqrt{\frac{|														    \mathcal{B}_\nu|}{2}}
												            \left(1+ 															    \sqrt{|			  												    \mathcal{D}_\nu|}														     \right)^{1/2}$\\
$\mathcal{B}_\nu<0$&						&					    & $\Omega_\nu=\pm i
													    \sqrt{\frac{|														    \mathcal{B}_\nu|}{2}}
												            \left(\sqrt{
													    |\mathcal 															    {D}|}-1\right)
													    ^{1/2}$\\
		& All roots are complex 			& All roots are real with		    & One pair of real
													      roots with \\			& in pairs of conjugates			& opposing signs			    &opposing signs
													     and one pair of\\
		& with opposing real parts			&		      	                    &imaginary roots   
													     with opposing 												     		     signs\\
\hline
		&$\Omega_\nu = \pm i \sqrt{\frac{|\mathcal{B}_\nu|}{2}} 
		  \left(1\pm i\sqrt{|\mathcal{D}_\nu|}\right)^{1/2}$&$\Omega_\nu=\pm i\sqrt{\frac{|\mathcal{B}_\nu|}{2}} 
								\left(1\pm \sqrt{|\mathcal{D}_\nu|}\right)
												     ^{1/2}$&$\Omega_\nu=\pm i
													      \sqrt{\frac{|														      \mathcal{B}_\nu|}														       {2}}\left(1+ 															\sqrt{|			  												    	\mathcal{D}_\nu|}														      \right)
													      ^{1/2}$\\
$\mathcal{B}_\nu>0$ & 						&					    &$\Omega_\nu=\pm 
													      \mbox{ }
													      \sqrt{\frac{|														      \mathcal{B}_\nu|}														       {2}}\left( 															\sqrt{|			  												    	\mathcal{D}_\nu|}														      -1 \right)
													      ^{1/2}$\\	
		& All roots are complex 			& All roots are imaginary with		    & One pair of real
													      roots with \\			& in pairs of conjugates			& opposing signs			    &opposing signs
													     and one pair of\\
		& with opposing real parts			&		      	                    &imaginary roots   
													     with opposing 												     		     signs\\
\hline \end{tabular}
\end{center}
\caption{Table of the eigenvalues $\Omega_\nu$ of the Jacobian $\mathcal{J}_\nu$. The table breaks the parameter space into $2$ regions (depicted as rows) demarcated by the sign of $\mathcal{B}_\nu$. In each zone, $\mathcal{D}_\nu$ further demarcates to $3$ sub-zones (depicted as columns). The eigenvalues, together with their natures, constitute the elements of the table.}
\label{tab:roots}
\end{table}
The eigenvalues with positive real parts depart further away from equilibrium in an unstable limit cycle. However, these trajectories are not expected to actually occur in these systems since they correspond to displacements that violate number conservation. In general, however, the radial fixed points have a saddle-like nature in this regime. In the case when $\Omega_\nu$ are purely real, then the radial fixed points remain saddles, but the decay to the radial orbit is non-spiral and very fast.

\begin{figure}[h!bt]
\ \epsfig{file=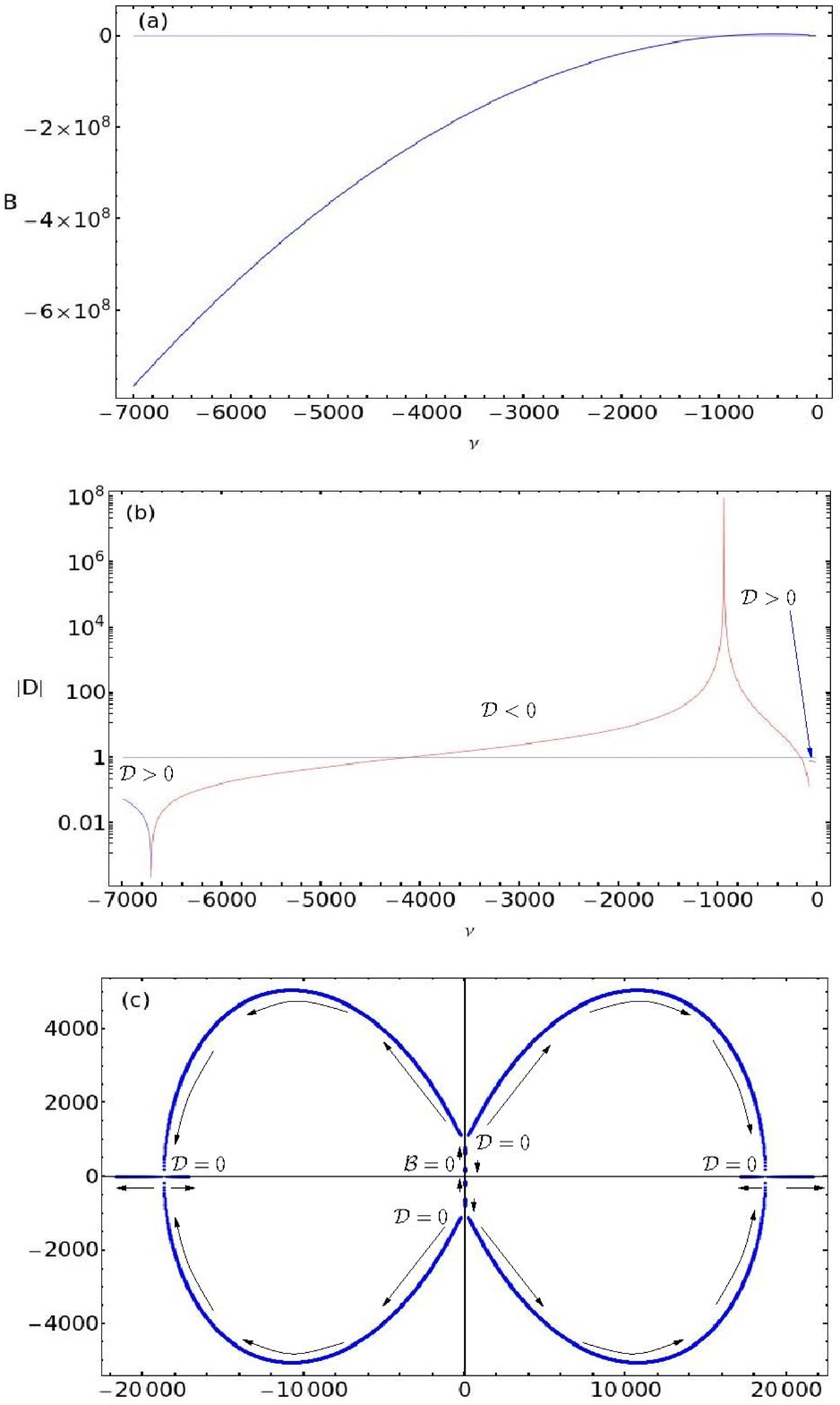,height=7.5in,width=4.0in}
\caption{(Color online) Bifurcation diagram in the complex plane of $\Omega_\nu$ obtained by numerically solving the BCS gap and number equations, applying the results to the Jacobian, and diagonalizing it. The numerics are performed for $100$ particles with $g_r=25$,$u_F = -0.3$,$u_B=0$, and renormalization cutoff $k_r=18.0$ in a unit volume for unit particle mass, where $\hbar$ has also been set to unity. Figure (a) shows the coefficient $\mathcal{B}$ defined in eq~\ref{nobosonint:coeff} as a function of detuning $\nu$. Note that $\mathcal{B} > 0$ until $\nu \approx -943.77(0)$ and then switches sign. Figure (b) shows the magnitude of the discriminant $\mathcal{D}$ as a function of detuning $\nu$ in a semi-log plot. The regions of positive (negative) $\mathcal{D}$ are indicated in blue (red). Figure (c) shows the actual bifurcation diagram, with the roots $\Omega$ shown in the complex plane. The origin, where $\mathcal{B}=0$ (see table~\ref{tab:roots}), is indicated, as well as the points where the discriminant $\mathcal{D}$ vanishes. The latter occurs at $\nu\approx-88.6(0)$ and $\nu\approx-6715.(0)$. The arrows indicate the direction where the magnitude of the detuning $\nu$ (given by $|\nu|$) increases in the range shown in figs (a) and (b).}
\label{fig:bif:nobosonint}
\end{figure}

 Note that the frequency $\Omega_\nu$ depends on the renormalization cutoff $k_R$ through $\alpha$, and so depends on the range of the Feshbach resonance induced interaction. For small ranges, $\Omega_\nu\sim k_R$ and so this differs from the oscillations about equilibrium seen in BEC systems with no Feshbach resonance. In general, a profile of the discriminant $\mathcal{D}_\nu$ and the coefficient $\mathcal{B}_\nu$ as functions of the detuning $\nu$, obtained using eqns~\ref{nobosonint:coeff} and~\ref{roots:jacobian:nobosonint}, will give us the bifurcation diagram of this system for any choice of parameters. Figure~\ref{fig:bif:nobosonint} (a) and (b) show plots of $\mathcal{B}_\nu$ and $\mathcal{D}_\nu$ as functions of detuning for the choice of parameters shown in the caption. Here, $\mathcal{B}_\nu$ is positive for small detuning and drops below zero at a critical value as $|\nu|$ is increased. Before that happens, however, $\mathcal{D}_\nu$ starts from a positive value that is less than unity. As table~\ref{tab:roots} indicates, this means that the eigenvalues are pure imaginary. While $\mathcal{B}_\nu$ is still positive, the discriminant drops to negative values at $\nu\approx--88.6(0)$, and the roots  are now complex conjugate pairs. The value of $\mathcal{B}_\nu$ drops down to zero at around $\nu=-943.77(0)$, with the roots now vanishing at the origin. As $\mathcal{B}_\nu$ becomes negative, $\mathcal{D}_\nu$ is also negative and the roots retain their ongoing structure until $\mathcal{D}_\nu$ switches back to positive at around $\nu\approx-6715.(0)$, after which the roots are real with opposite signs as per table~\ref{tab:roots}. All these trends can be seen in the numerical evaluation of the eigenvalues plotted in the bifurcation diagram in figure~\ref{fig:bif:nobosonint} (c). Thus, a bifurcation is seen at the locus of $\mathcal{D}_\nu=0$ in the shallow-BEC regime. Here the limit cycles coalesce into orbital trajectories around the fixed points as the detuning $\nu$ is varied. However, this is not a Hopf bifurcation. The real parts of the eigenvalues do not cross over from the negative to the positive side as would be the case with a Hopf bifurcation~\cite{strogatz:book}. 

The paragraph above demonstrates that the 'deep-BEC ' and 'shallow-BEC' regimes can be differentiated by the sign of the discriminant $\mathcal{D}_\nu$ inside these regimes. Finally, a rapid quench from the former to the latter can be expected to yield orbital phase space dynamics of $\Psi_j$ around the equilibrium fixed points after a sufficiently long time. Variations in the BEC number density, given by $|\Psi_2|^2 = x^2_3 + x^2_4$, is thus expected to show rapid nontrivial oscillations that can qualitatively resemble some sort of 'collapse and revival' of the matter wave packet.

\begin{figure}[h!bt]
\hspace*{-0.3in}
\ \epsfig{file=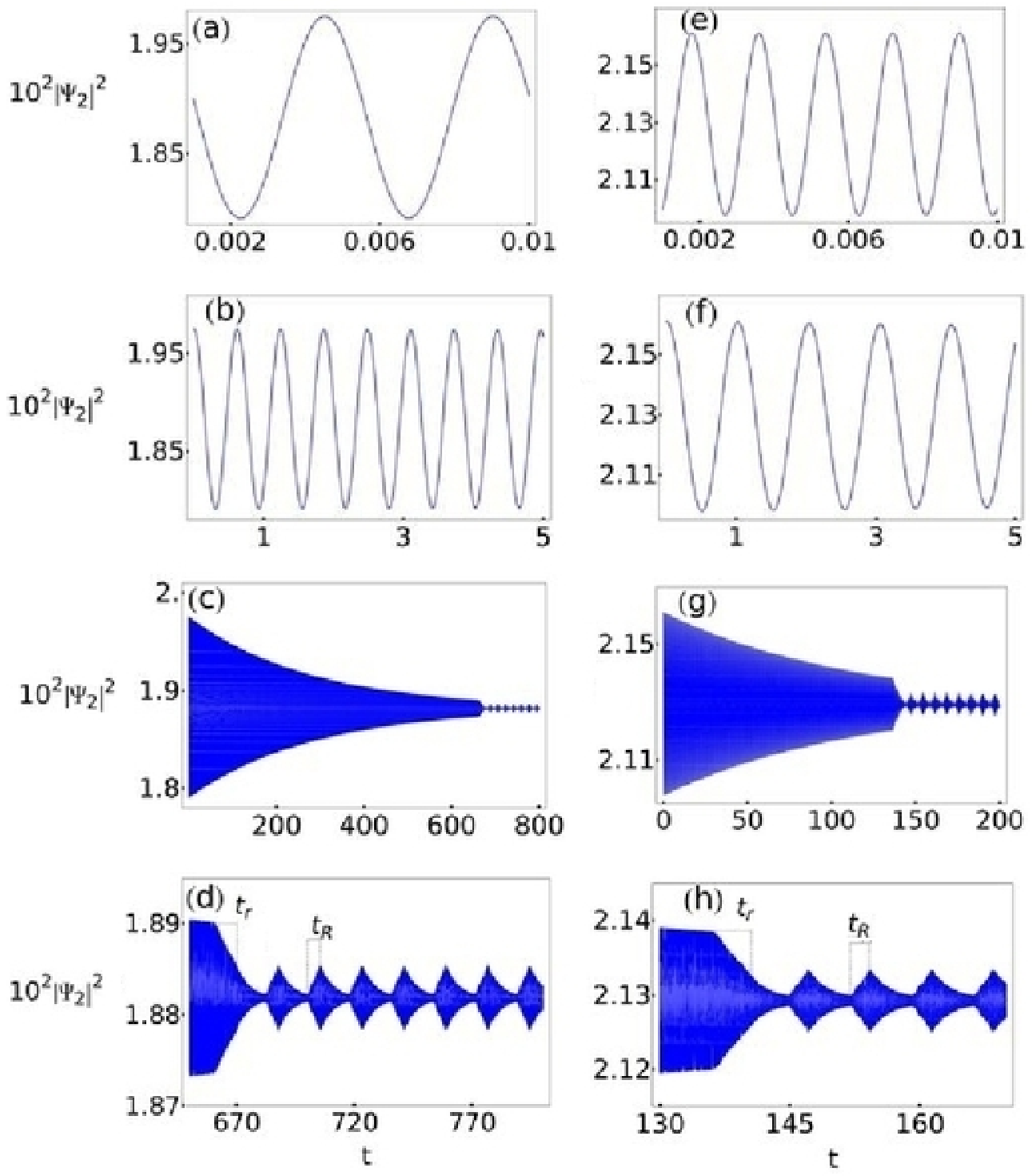,height=6.2in,width=6.2in}
\caption{(Color online) Plots of the signal $|\Psi_2|^2(t)$  from the solutions to equations~\ref{dynamical_system} for $u_B = 0$ after the system is quenched from a pure BEC to the shallow-BEC regime (see figure~\ref{fig:chempot}). Here, $\hbar = m =\mathcal{V} = 1$, and $u_F = -0.3$. Figures (a) - (d) show plots of the time evolution for $g_r=25.0$ and $\nu = -55.0$. Figures (e) - (h) show plots for $g_r=40.0$ and $\nu = -140.0$. Figures (a) and (e) show plots for very small times ($\sim 10^{-2}$). Figures (b),(c) and (f),(g) show plots of discrete time samples of the signal taken in units of the local time period from figures (a) and (e) respectively. Thus, figures (b),(c) and (f),(g) show the signals that envelope the master signal in (a) and (e) respectively. For figure (a), the master signal has frequency $\Omega_m \simeq 1400$, and for figure (e) $\Omega_m \simeq 3500$. Figures (b),(c) and (f),(g) show the same corresponding envelope plots for two different time intervals. Partial collapse and revival of the matter wave can be seen after a lengthy decay time, where the original envelope damps out as it oscillates. The envelope oscillation frequency $\Omega_e \simeq 10$ for figure (b),(c) and $\simeq 6$ for figures (f),(g). The decay times $t_d$ for figures (c) and (g) are approximately $400$ and $125$ respectively. The region where collapse and revival take place is magnified and shown in figures (d) and (h), where the entire signal (not just the envelope) is now being plotted. The presence of partial collapse and revival of the matter wave can be noted after a relaxation time $t_r$ as indicated in figures (d) and (h). The revival times $t_R$ are also indicated in these figures. In figure (d), $t_r\simeq9$, $t_R\simeq6$. In figure (h), $t_r\simeq4$, $t_R\simeq2.5$}
\label{fig:colrev}
\end{figure}

Now, the dynamics is analyzed numerically for larger values of $\epsilon_F$ in the shallow-BEC regime. In this region, numerical methods are utilized to solve equations~\ref{dynamical_system}. The system is evolved numerically using Runge-Kutta-Fehlberg methods using the initial conditions $\Psi_{1,2} = g^2_r / 2 \epsilon_F$, corresponding to $|b_0|^2 = \mathcal{N}/2$ with $\mathcal{N}$ taken to be $100$. The quenching is assumed to have taken place by a sharp variation of $\epsilon_F$ (via the Feshbach detuning $\nu$) from a large negative value (in the deep-BEC regime) to the shallow-BEC region. Figures~\ref{fig:colrev}(a)-(h) show the time variation of the BEC condensate fraction for the two representative parameter sets from figure~\ref{fig:chempot} in section~\ref{sec:fpchempot}. Figures~\ref{fig:colrev}(a) and (e) show the time evolution for very small times ($ \sim 10^{-2}$ units). These signals are the master signals of the dynamics. The master signals appear to be Rabi oscillations about the equilibrium value $|\bar{\Psi}_2|^2$ with a frequency of $\Omega_m$. In figures~\ref{fig:colrev}(b) and (f), discrete time samples of the actual signals are plotted in steps of $2\pi/\Omega_m$. The presence of another oscillation(s) of frequency $\Omega_e$ indicates that the actual signals contain a superposition of several comparably fast Rabi oscillations interfering with each other, and that $\Omega_e$ is their \textit{beat frequency}. For larger times, note from figures~\ref{fig:colrev}(c) and (g) that these beats damp out over a long period of time $t_d$, until a sudden onset of partial \textit{collapse and revival} of the matter wave begins after a short initial relaxation time $t_r$ (indicated in the respective figures). The \textit{revival time} $t_R$ is also indicated in the respective figures. Approximate values of $t_d$, $t_r$, $t_R$ and $\Omega_{m,e}$ have been obtained from figures~\ref{fig:colrev} and can be seen in table~\ref{tab:colrev}. 
The numerics plotted in fig~\ref{fig:colrev} indicate that the onset of this collapse and revival takes place at around $10^3$ numerical units. The values of $\epsilon_F$ , which are of the order of $10^2$ in numerical units, can be obtained from table~\ref{tab:colrev}. Note that these times are expressed in numerical units. In order to obtain them in actual units, it is necessary to start with the system trap size $L_u$, and note that the Fermion mass ($M$ in actual units, $m$ in numerical units) is taken to be unity in these numerical units. Thus, the relationship between numerical time $t$ and actual time $T$ can be obtained using eqns~\ref{greek_consts}, ~\ref{greek_consts_chempot}, the dimensionless scaling of time by $(\epsilon_F d)^{-1}$ in equations~\ref{dynamical_system} , and the fact that the base unit of time is $T_u = 2 M L^2_u/\hbar$. The relationship is
\begin{equation}
 L_u=2\left(\sqrt{\frac{2}{\epsilon_F}}\frac{\pi \hbar }{M}\frac{T}{t}\right)^{1/2},
\end{equation}
where the chemical potential $\epsilon_F$ is expressed in numerical units. Thus, the trap size $L_u$ can be adjusted for a particular $t$ until an experimentally accessible $T$ is attained. As an example, if the atoms involved are Fermionic isotopes of Lithium (with mass $M\sim 10^{-26}$ kg), and the onset of collapse and revival (at around $t\sim 10^3$) is desired to be of the order of milliseconds, then the equation above necessitates that the trap size be of the order of $300$ nanometers, with the larger confinements requiring longer times $T$ growing as  $\sqrt{T}$. Ultracold BCS systems with such dimensions have already been obtained using superconducting atom chips~\cite{atom:chips}, where trap lifetimes can range from seconds to minutes~\cite{atomchips:traps}. Therefore, such time scales are experimentally accessible. The similarities in order of magnitude with $\Omega_m$ from table~\ref{tab:colrev} and $\Omega_{b_0}$ from figure~\ref{fig:omega} suggest that the collapse and revival effect may be soliton-like Rabi oscillations reported in the literature. Also, note from these figures that all three time scales, $t_d$, $t_r$ and $t_R$, are larger for smaller $g_r$. Thus, it is expected that $t_d \rightarrow \infty$ as $g_r \rightarrow 0$, removing the collapse and revival effect. The master signal frequency is also expected to reduce to $2\lambda_\nu$ in this limit. It is also noted from figure~\ref{fig:chempot}(d) that the revival process characterized by $t_R$ seems to be interrupted by a signal that has the same temporal characteristics as the relaxation signal characterized by $t_r$, thereby inhibiting a total revival of the matter wave. 

In the case of interacting Bosons, the parameters from figure~\ref{fig:chempot2} are used and  the coefficient $\beta$ neglected to simplify the dynamics. Thus, equations~\ref{dynamical_system} simplify to
\begin{eqnarray}
\label{dynamical_system_nobeta_int}
\dot \Psi_1 + i\gamma \left( \Psi_1-\Psi_2 \right) - i\alpha \Psi_1  &=& 0 \nonumber \\
\dot \Psi_2 +2 i \lambda \Psi_2 +2 i\chi|\Psi_2|^2\Psi_2 - i\kappa\gamma \left(\Psi_1-\Psi_2 \right) &=& 0,
\end{eqnarray}
This system is evolved numerically using the same initial conditions and numerical methods as detailed in the previous paragraph (however, the tolerance required for convergence was an order of magnitude lower). The Boson interaction amplitude was taken to be fairly small so as to see the effects of departing slightly from the $u_B=0$ case for equations~\ref{dynamical_system_nobeta}. For small values of $u_B$, only a damped oscillation was seen for all the time intervals that were allowed by the numerical tolerances. For $u_B = 0.6$, the collapse and revival effect was seen for large times, just as it was with the case of noninteracting bosons. 

\begin{figure}[h!bt]
\hspace*{-0.3in}
\ \epsfig{file=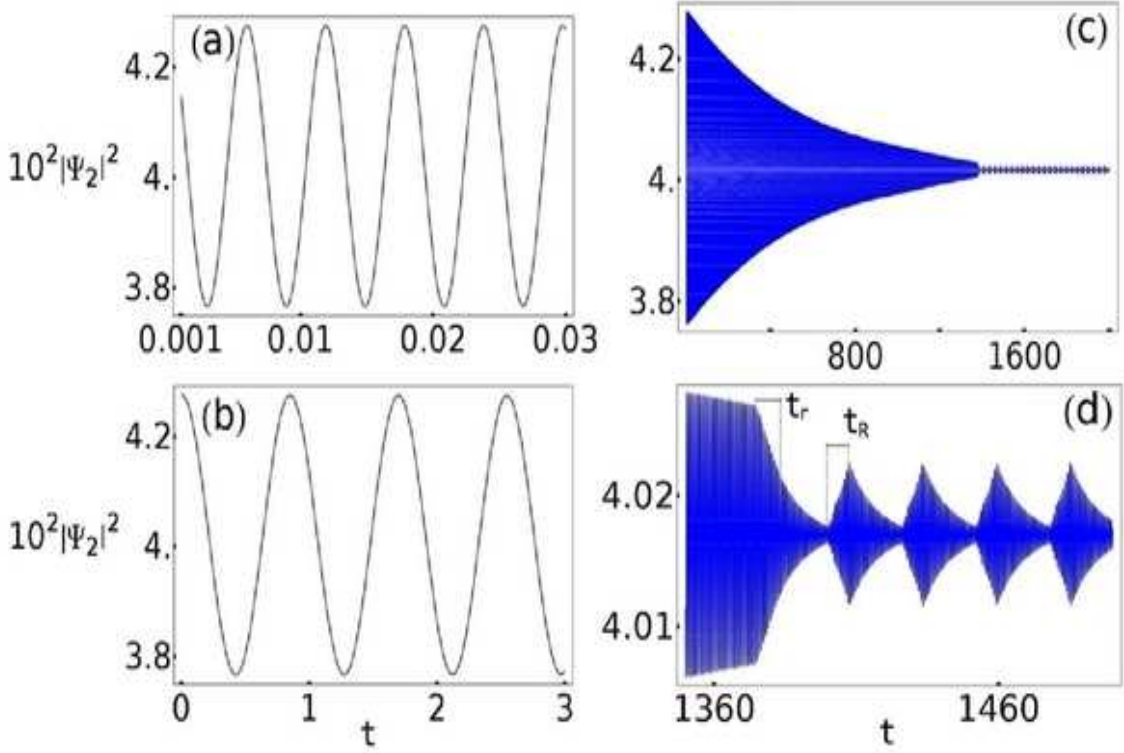,height=4in,width=6in}
\caption{(Color online) Plots of the signal $|\Psi_2|^2(t)$  from the solutions to equations~\ref{dynamical_system_nobeta} for $u_B = 0.6$ after the system is quenched from a pure BEC to the shallow-BEC regime (see figure~\ref{fig:chempot}). Here, $\hbar = m =\mathcal{V} = 1$, $u_F = -0.3$, and $g_r = 25$.Figures (a) shows plots for very small times ($\sim 10^{-2}$). Figure (b) shows plots of discrete time samples of the signal taken in units of the local time period from figure (a) . Thus, figure (b) shows the signal that envelopes the master signal in (a). For figure (a), the master signal has frequency $\Omega_m \simeq 1050$. Figures (b) and (c) show the same corresponding envelope plots for two different time intervals. Partial collapse and revival of the matter wave can be seen after a lengthy decay time, where the original envelope damps out as it oscillates. The envelope oscillation frequency $\Omega_e \simeq 1$. The region where collapse and revival take place is magnified and shown in figure (d). The presence of partial collapse and revival of the matter wave can be noted after a relaxation time $t_r$ as indicated in figure (d). The revival times $t_R$ are also indicated in these figures. In figure (d), $t_r\simeq 12$, $t_R\simeq 8$.}
\label{fig:colrev2}
\end{figure}

In both of these cases, the nonlinear contribution to the dynamics go as $|x|^2 x$, absent which the dynamics is oscillatory. Therefore, it is believed that this effect is caused by the nonlinear dynamics from this type of expression alone. The results of the numerical simulation for $u_B = 0.6$ are shown in figures~\ref{fig:colrev2}(a) through (d). The frequencies $\Omega_m$  and $\Omega_e$, as well as the time scales $t_d$, $t_r$ and $t_R$, are tabulated in table~\ref{tab:colrev}. Note that the decay time $t_d$ is much higher for the case of interacting bosons and no $\beta$, suggesting that the contribution of the Boson interaction to the collapse and revival is much smaller than that of the term that is nonlinear in the gap parameter and weighed by $\beta$. These striking phenomena should manifest in the time-of-flight absorption images of the gas.
\begin{table}
\begin{center}
\begin{tabular}{ | l  l  l  l|| l | l | l | l | l |}
      \hline
  $g_r$ 	& $-\nu$  & $\epsilon_F$ &$u_B$	& $\Omega_m$    & $\Omega_e$     & $t_d$ 	& $t_r$ & $t_R$ \\  \hline  \hline
  $25$		& $55$	  & $125.78$ 	 &$0$ 	& $1400$	& $10$		 & $400$ 	& $9$	& $6.0$   \\  \hline
  $40$		& $140$	  & $192.40$ 	 &$0$	& $3500$	& $6$		 & $125$	& $4$	& $2.5$ \\  \hline
  $25$		& $110$	  & $85.5$	 &$0.6$ & $1050$	& $1$		 & $600$	& $12$  & $8.0$ \\  \hline
     \end{tabular}
\caption{Table of master and envelope frequencies $\Omega_{m,e}$, as well as decay, relaxation and revival times ($t_{d,r,R}$ respectively) of the quenched dynamics shown in figures~\ref{fig:colrev} and~\ref{fig:colrev2}. The system is evolved for two values of $g_r$. The corresponding $\nu$s and $\epsilon_F$s are also shown.  {The values of the parameters are chosen as a representative sample, based on our choice of length and atom number scales as explained in section}~\ref{sec:fpchempot},  {and the conclusions are not affected qualitatively if the ranges are varied within these orders of magnitude and kept in the regions of interest.}}
\label{tab:colrev}
\end{center}
\end{table}

\section{Conclusion}
\label{sec:conclusion}
In this paper, the dynamics of an ultracold gas of Fermions with a narrow Feshbach resonance has been formulated in the dual-channel case after making the single mode approximation for Bosons. The mean field nonlinear and complex TDGL dynamics of the Fermions, coupled to the Gross Pitaevski dynamics for the composite Bosons, have been obtained. The Fermion dynamics is encapsulated in the temporal variation of the superfluid gap parameter $\Delta$, and that of the Bosons in the ground state amplitude $b_0$.

The equilibrium states for the system in the BEC side have been evaluated by looking at the adiabatic evolution of the stationary solutions of the dynamics, which leads to a population transfer from the BEC superfluid to the BCS superfluid for nondivergent Fermion interactions. The dynamics of the system as it is quenched from a pure BEC state to a state in the shallow-BEC  regime (accomplished by a rapid variation in the Feshbach detuning $\nu$) has been analyzed numerically. Nonlinearities in the dynamics cause the Rabi oscillations to relax back to it's equilibrium state. However, at large times, the relaxation gets interrupted and a
\textit{collapse and revival} type phenomenon of the Bosonic matter wave field ensues. This effect seems to be caused by the interference between the multiple modes of the dynamics, and shows a striking phenomenon that is analogous to the collapse and revival effect that has been reported experimentally for Bosons. 
\section{Acknowledgements}
This paper was supported by a postdoctoral fellowship from the Department of Science and Technology, Government of India,  at the SN Bose National Centre for Basic Sciences. The author thanks both institutions for financing this work. The author also thanks Prof J.K. Bhattacharjee for his discussions and insights into the problem.
\pagebreak
\pagebreak

\end{document}